\documentclass[prd,preprint,floatfix,nofootinbib,tightenlines]{revtex4-2}

\usepackage[english]{babel}
\usepackage{amsmath,amssymb,bm,slashed}
\usepackage{graphicx}
\usepackage[sort&compress]{natbib}
\usepackage[usenames,dvipsnames,svgnames]{xcolor}
\usepackage[normalem]{ulem}
\usepackage{hyperref}
\usepackage[capitalize]{cleveref}
\usepackage{subfigure} 
\definecolor{red}{rgb}{1.0, 0, 0}
\definecolor{green}{rgb}{0.0,0.7,0.2}

\hypersetup{
  colorlinks,
  linktoc=page,
  linkcolor=blue,
  citecolor=ForestGreen
}

\allowdisplaybreaks

\bibpunct{[}{]}{,}{n}{}{,}

\newcommand{\ev}[1]{\ensuremath{\left\langle #1 %
                     \right\rangle}} 
\newcommand{\tr}{\text{tr}}

\renewcommand{\vec}[1]{{\mathbf{#1}}}

\newcommand{\beq}{\begin{equation}}
\newcommand{\eeq}{\end{equation}}

\newcommand{\pr}{\ensuremath{\mbox{pr}}}
\DeclareMathOperator{\Normal}{Normal}

\newcommand{\astar}{\ensuremath{\mathbf{a}^\star}}
\newcommand{\atrue}{\ensuremath{\hat{\mathbf{a}}}}
\newcommand{\En}{\mathbb{E}_N}
\newcommand{\Ez}{\mathbb{E}_z}
\newcommand{\avg}[1]{\ensuremath{\left\langle #1 \right\rangle}}
\newcommand{\avgn}[1]{\avg{#1}_N}
\newcommand{\avgz}[1]{\avg{#1}_z}
\newcommand{\LL}{\log L}

\usepackage{amsthm}

\newtheorem{theorem}{Theorem}

\begin{document}

\title{Bayesian model averaging for analysis of lattice field theory results}
\author{William I.~Jay$^{1}$}		 \email[Email: ]{wjay@fnal.gov}
\author{Ethan T.~Neil$^{2}$}              \email[Email: ]{ethan.neil@colorado.edu}
\affiliation{$^1$ Theoretical Physics Department, Fermi National Accelerator Laboratory, Batavia, IL 60510, USA}
\affiliation{$^2$ Department of Physics, University of Colorado, Boulder, CO 80309, USA}
\date{August 3, 2020} 

\begin{abstract}

Statistical modeling is a key component in the extraction of physical results from lattice field theory calculations.
Although the general models used are often strongly motivated by physics,  many model variations can frequently be considered for the same lattice data.
Model averaging, which amounts to a probability-weighted average over all model variations, can incorporate systematic errors associated with model choice without being overly conservative.
We discuss the framework of model averaging from the perspective of Bayesian statistics, and give useful formulae and approximations for the particular case of least-squares fitting, commonly used in modeling lattice results.
In addition, we frame the common problem of data subset selection (e.g. choice of minimum and maximum time separation for fitting a two-point correlation function) as a model selection problem and study model averaging as a straightforward alternative to manual selection of fit ranges.
Numerical examples involving both mock and real lattice data are given.

\end{abstract}

\begin{flushright}
FERMILAB-PUB-20-374-T
\end{flushright}

\maketitle

\newpage
\section{Introduction}
\label{sec:intro}

One of the central problems in lattice field theory is that of model fitting and parameter estimation.  This problem appears repeatedly in analysis of lattice results, from single two-point correlators up to joint chiral and continuum extrapolations of results from many simulation streams.  The functional forms which appear in these cases are often notoriously difficult to work with; the sum of exponentials which models the two-point correlator is in general quite numerically unstable, and chiral and continuum extrapolations can involve nonlinear dependence on large numbers of unknown variables.

Making analysis of lattice simulations even more challenging, many of the models appearing require an ever-increasing number of parameters as simulations become more precise.  For example, chiral perturbation theory is an effective theory which will break down at large masses or momentum scales, and contains (in principle) an infinite number of low-energy constants.
The spectral decomposition of a single 
two-point correlator also contains (in principle) an infinite tower of excited states.
The Symanzik effective theory describing the appearance of lattice artifacts similarly contains an organized but infinite number of terms.
These contributions are typically dealt with by truncating the model, and often the data as well.  
However, this can lead to subtle dependence in the results on the analyst's choice of fit range and number of terms included in the model.

This is not to say that lattice theorists are unaware of these potential sources of systematic error.  The effects of model truncation and data truncation can be estimated by studying the variation of quantities of interest as the range of data included is varied, or additional model terms are added.  However, the often-adopted approach of taking the full difference between these variations as a systematic error is somewhat crude and likely to be overly conservative in many cases.

In this paper, we describe the technique of Bayesian model averaging as an alternative approach to these systematic errors, and outline its potential applications in the analysis of lattice simulation results.  This approach allows for fully rigorous estimation of probability distributions for parameters of interest by combining results from several models.  All models must reduce to a common model containing the parameter(s) to be estimated, but there is no other requirement that they be nested models.  (For a continuum extrapolation of a matrix element e.g., the reduced model can be simply be a single constant parameter, the value of the matrix element in the continuum limit.)

Bayesian model averaging is somewhat well-known in the statistical literature \cite{leamer1978specification,racine1986bayesian,george1993variable,kass1995bayes,raftery2003discussion}, although it is most often used in the context of linear models.
Here we place no such restriction, giving formulae which can be used for arbitrary nonlinear models.\footnote{
Appendix~\ref{sec:linear_vs_nonlinear} defines the distinction between linear and nonlinear models.
}
In general, the model probability weights required for model averaging are complicated integrals, but we will give approximate formulae which may be used with sufficient sample sizes.  
As we will show, combining statistical results obtained with a set of models $\{M\}$ relies on the estimation of the model weight, $\pr(M|D)$.
We will also argue that the commonly-used procedure of applying cuts to the data can also be understood as a model selection problem.
Our key result is that, at leading order in the data sample size $N$, the model weight can be estimated by the Akaike information criterion (or ``AIC''), modified with a penalty term for cutting away data points:
\beq
\pr(M|D) \approx \exp \left[ -\frac{1}{2} \left( \chi_{\rm aug}^2(\mathbf{a}^\star) + 2k + 2N_{\rm cut} \right) \right]
\eeq
(this is \cref{eq:model_avg_cut}, simplified for the case of uniform model priors).
Here $\chi_{\rm aug}^2(\mathbf{a}^\star)$ is the standard best-fit augmented chi-squared \cite{Lepage:2001ym} (see \cref{eq:chi2_aug} below), $k$ is the number of fit parameters, and $N_{\rm cut}$ is the number of removed data points as defined in \cref{sec:subset}.

Our work is partially inspired by the Bayesian approach to effective field theory advocated by Schindler and Phillips \cite{Schindler:2008fh}.  Other examples of using weighted averages over models and Bayesian ideas in lattice analysis include \cite{Davies:2008sw, Durr:2013goa, Berkowitz:2017gql, Chang:2018uxx, Rinaldi:2019thf, Miller:2020xhy,Borsanyi:2020mff,Chen:2004gp}.  We believe that our treatment is the first attempt to lay out the procedure rigorously and in a fully Bayesian framework for a lattice field theory audience.

An outline of the paper is as follows.
In \cref{sec:framework}, we describe the basic Bayesian framework for model averaging and derive formal results for model-averaged expectation values.
\cref{sec:lsq} specializes to the case of least-squares fitting and derives an approximate formula for the model probabilities which are needed for model averaging.  \cref{sec:ex1} gives a numerical example of model averaging applied to mock data.
In \cref{sec:subset}, we discuss the common problem of data subset selection, reframing it as a model variation problem in order to apply model averaging.
\cref{sec:ex2} shows an example application to the common task of fitting a two-point correlation function and demonstrates the effectiveness of model averaging as a replacement for choosing cuts on the data.
We make some concluding remarks in \cref{sec:conc}.
A detailed discussion of bias correction in the estimation of model probability is given in \cref{sec:bias-appendix}.  \cref{sec:chi-squared-appendix} and \cref{sec:linear_vs_nonlinear} describe some technical details related to the chi-squared function and linear vs. nonlinear models.

\section{Bayesian framework}
\label{sec:framework}

The basic analysis problem is as follows: we wish to describe a dataset $D$ using a base model $M_0$ in order to determine the value of one or more common parameters $\{\mathbf{a}_{c}\}$.
In the example of a continuum extrapolation, $M_0$ could simply be the value of a specific matrix element in the continuum limit.

In estimating the parameters $\{\mathbf{a}_{c}\}$, we are often led to consider several extensions of the base model $M_0$ which contain various unphysical or uninteresting terms, like lattice artifacts or undetermined excited states.
This extends our study to a space of models $\{M\}$, but our basic interest is still in estimation of the parameters of $M_0$.
All of the models in $\{M\}$ must contain $M_0$, in the sense that marginalizing over additional parameters $\{\mathbf{a}_m\}$ reduces them to $M_0$. (Note that the set $\{\mathbf{a}_m\}$ implicitly depends on the choice of model $M$.)

It is important to note that the base model $M_0$ itself does not necessarily have to be contained in the set $\{M\}$ of models that  are actually fit to the data.
As a simple example, in a continuum extrapolation it is certainly not necessary to include the continuum-only model $M_0$ (without any lattice spacing dependence) in the set of fits.
Indeed, a continuum-only model would surely describe the data poorly in this example.

To obtain the marginal probabilities for the common parameters, we marginalize over both models and additional parameters \cite{Schindler:2008fh}:
\beq \label{eq:master}
\pr(\mathbf{a}_{c} | D) = \sum_M \int d\mathbf{a}_{m} \frac{\pr(D|\mathbf{a},M) \pr(\mathbf{a}|M) \pr(M)}{\pr(D)}.
\eeq
where ``$\pr$'' denotes a probability distribution, and the set of all parameters $\{\mathbf{a}\}$ is the union of $\{ \mathbf{a}_c\}$ and $\{ \mathbf{a}_m \}$.
In principle, this formula assumes that all parameters $\{\mathbf{a}\}$ are dimensionless.
In practice, we will be interested in model weights which will depend only on ratios of probabilities, so any units will tend to cancel.

If we can carry out the integrals and explicitly construct $\pr(\mathbf{a}_c | D)$, then expectation values for arbitrary functions of the common parameters $\mathbf{a}_c$ are immediately available:
\beq
\ev{f(\textbf{a}_c)} = \int d\textbf{a}_c f(\textbf{a}_c) \pr (\textbf{a}_c | D),
\eeq
from which we can construct the standard mean, variance, and so forth.
However, evaluating the integrals in the ``master formula'' \cref{eq:master} is generally quite difficult, especially in the context of the complicated non-linear models appearing in lattice analyses.\footnote{Direct Monte Carlo evaluation of such integrals is an intriguing option which deserves more attention in the context of lattice studies, in which much more complicated integrals are evaluated as a matter of course.
This method seems to have been explored in Ref.~\cite{Morningstar:2001je}.  However, we will not pursue this approach here.}

For our present purposes, it is more interesting to observe that we can reconstruct the combined estimate from the individual model fit results.
Applying Bayes' theorem and using elementary properties of conditional probability gives
\begin{align}
\ev{f(\mathbf{a}_{c})}_{M} &= \int d\mathbf{a}_{c} f(\mathbf{a}_{c}) \pr(\mathbf{a}_{c} | M,D) \\
&= \int d\mathbf{a}_{c} f(\mathbf{a}_{c}) \frac{\pr(\mathbf{a}_{c}, M, D)}{\pr(M, D)} \\
&= \int d\mathbf{a}_{c} f(\mathbf{a}_{c}) \frac{\pr(D|\mathbf{a}_{c}, M) \pr(\mathbf{a}_{c}, M)}{\pr(M|D) \pr(D)} \\
&= \int d\mathbf{a}_{c} f(\mathbf{a}_{c}) \frac{\pr(D|\mathbf{a}_{c}, M) \pr(\mathbf{a}_{c} | M) \pr(M)}{\pr(M|D) \pr(D)} \\
&= \frac{1}{\pr(M|D)} \int d\mathbf{a}_{c} f(\mathbf{a}_{c}) \pr(\mathbf{a}_{c}, M | D).
\end{align}
But now if we marginalize the integral on the right-hand side over the space of models $\{M\}$, we just obtain the total, model-independent expectation value for $f$:
\beq
\sum_M \int d\mathbf{a}_{c} f(\mathbf{a}_{c}) \pr(\mathbf{a}_{c}, M | D) = \int d\mathbf{a}_{c} f(\mathbf{a}_{c}) \pr(\mathbf{a}_{c} | D) = \ev{f(\mathbf{a}_{c})}.
\eeq
Thus, we arrive at the relation
\beq
\ev{f(\mathbf{a}_c)} = \sum_M \ev{f(\mathbf{a}_c)}_M \pr(M|D)\label{eq:model_avg}.
\eeq
This is the central formula of interest for purposes of model averaging.  It shows that all of the moments of the fully combined probability distribution function can be obtained as a weighted average over individual model information, with the weight factors given by the posterior probability $\pr(M|D)$ for each individual model.  
Due to its role in model averaging, we will refer to $\pr(M|D)$ interchangeably as the ``posterior probability'' or as the ``model weight."
The model weight itself can be expressed as an integral over the parameter space,
\begin{align}
\pr(M|D)
    &= \int d\vec{a}\ \pr(M,\vec{a}|D) \\
    &= \int d\vec{a}\ \frac{\pr(D|\vec{a},M) \pr(\vec{a},M)}{\pr(D)} \\
    &= \int d\vec{a}\ \frac{\pr(D|\vec{a},M) \pr(\vec{a}|M) \pr(M)}{\pr(D)}.\label{eq:model_weight}
\end{align}
These probabilities are normalized to unity,
\beq
\sum_M \pr(M|D) = 1,
\eeq
which follows immediately from the definition of conditional probabilities and the marginalization formula, $\sum_M \pr(M,D) = \pr(D)$.
Another (but perhaps more physical) argument to obtain the same result is that the expectation of the unit operator $\ev{1}$ should be unity independent of model choice.

The formulae presented so far are completely general.
However, certain common choices used for the estimators of the likelihood function and other quantities can introduce bias.
Any such biases should be corrected to guarantee convergence to correct results.
We discuss an important bias correction in detail in \cref{sec:bias} below.

\subsection{Estimation of model parameters with model averaging \label{sec:ma_formulas}}

It is instructive to consider what happens to the simple estimate of a model parameter under the model combination procedure.
Suppose we are interested in the single parameter $a_0$, marginalized over a set of $N_M$ models $\{M\}$.
Using \cref{eq:model_avg}, we find for its mean
\beq
\ev{a_0} = \sum_M \ev{a_0}_M \pr(M|D) \label{eq:model_avg_mean}
\eeq
and variance:
\begin{align}
\sigma_{a_0}^2 &= \ev{a_0^2} - \ev{a_0}^2 \\
&= \sum_{i=1}^{N_M} \ev{a_0^2}_i \pr(M_i | D) - \left(\sum_{i=1}^{N_M} \ev{a_0}_i  \pr(M_i | D) \right)^2 \\
&= \sum_{i=1}^{N_M} \sigma_{a_0, i}^2 \pr(M_i | D) + \sum_{i=1}^{N_M} \ev{a_0}_i^2 \pr(M_i | D) - \left(\sum_{i=1}^{N_M}  \ev{a_0}_i \ \pr(M_i | D) \right)^2.\label{eq:model_avg_variance}
\end{align}
This result for the variance also appears in the statistics literature \cite{kass1995bayes}, and has been used in the context of lattice calculations in \cite{Chang:2018uxx,Miller:2020xhy}. 
The first term is simply the weighted average of the statistical variance over all models.
The remaining two terms can then be thought of as a ``systematic error" contribution to the variance of $a_0$ due to model choice.
In the special case of equal model weights, i.e., $\pr(M_i | D) = 1/N_M$, the latter contribution can be thought of as the variance over the space of models, since it reduces to the standard formula
\beq
\sigma_{a_0, \textrm{syst}}^2 = \frac{1}{N_M} \sum_{i=1}^{N_M} \ev{a_0}_i^2 - \frac{1}{N_M^2} \left( \sum_{i=1}^{N_M} \ev{a_0}_i \right)^2.
\eeq
We note that in the general case, this is \textit{not} the same as the variance computed from the set of weighted estimates $w_i \equiv \ev{a_0}_i \pr(M_i | D)$; such a weighted variance would contain an extraneous factor of $\pr(M_i | D)$ in the $\ev{a_0}^2$ term.  We also note that taking the full width of the distribution of results $\ev{a_0}_i$, as done in e.g.\ \cite{Bazavov:2014wgs}, will give an estimated systematic error strictly greater than $\sigma_{a_0, \rm{syst}}$, so that this procedure is a conservative error estimate.

It is illustrative to specialize to the case of considering only two models $M_1, M_2$, for which we have found the weights
\begin{align}
\pr (M_1 | D) &= 1-p, \\
\pr(M_2 | D) &= p.
\end{align}
Suppose now that model 1 is strongly favored by the data, so $p$ is small.  Expanding the expectation values above to first order in $p$, we find:
\begin{align}
\ev{a_0} &= \ev{a_0}_1 + (\ev{a_0}_2 - \ev{a_0}_1) p, \\
\sigma_{a_0}^2 &\approx \sigma_{a_0,1}^2 + \left[ \sigma_{a_0,2}^2 - \sigma_{a_0,1}^2 + (\ev{a_0}_2 - \ev{a_0}_1)^2 \right] p.
\end{align}
In the limit $p \rightarrow 0$ we recover the statistical results of $M_1$, as expected.  For small but non-zero $p$, the corrections to the mean and variance of $a_0$ due to including $M_2$ are \emph{likely} to be small, but it is clear that this depends on how large the difference between the estimates from $M_1$ and $M_2$ are.

\section{Least-squares fitting}
\label{sec:lsq}

The discussion so far has been completely general with regards to the form of the probability distributions appearing.
We now specialize to the most common usage case in the context of lattice simulations, namely least-squares regression of a model $M$ to some data set $D$.
The likelihood function $\pr(D | \mathbf{a}, M)$ is taken to be
\beq \label{eq:Lsample}
\pr(D|\mathbf{a}, M) = \prod_{i=1}^{N} \frac{1}{(2\pi)^{d/2} (\det \Sigma)^{1/2}} \exp \left[ -\frac{1}{2} \chi_i^2 \right],
\eeq
where
\beq \label{eq:chi2sample}
\chi_i^2 \equiv (y_i - f_M(\mathbf{a}))^T \Sigma^{-1} (y_i - f_M(\mathbf{a})) 
\eeq
is the standard chi-square goodness of fit\footnote{
This form is standard in the statistics literature.
For practical applications, another definition based on the sample means is generally used, $\chi^2 = (\bar{y} - f_M(\mathbf{a}))^T \Sigma^{-1} (\bar{y} - f_M(\mathbf{a}))$.  Up to data-dependent constants, these two definitions are actually identical; see \cref{sec:chi-squared-appendix} for a derivation.
}
involving the data sample $y_i$ and the model function $f_M(\mathbf{a})$; we assume the samples are drawn independently from some underlying distribution.
Here $d$ denotes the dimension of a single observation vector $y_i$, and $N$ is the number of independent observations drawn from the true distribution.
The matrix
$\Sigma = \frac{1}{N-1} \sum_{i=1}^N (y_i - \bar{y}) (y_i - \bar{y})^T$
is the covariance matrix between the $y_i$.

For the prior distribution, it is standard (and typically justified by the principle of maximum entropy \cite{Lepage:2001ym,Schindler:2008fh}) to adopt a multivariate Gaussian form,
\begin{align}
\pr(\mathbf{a} | M)
&= \frac{1}{(2\pi)^{k/2} (\det \tilde{\Sigma} )^{1/2}} \exp \left[ -\frac{1}{2} (\mathbf{a} - \tilde{\mathbf{a}})^T \tilde{\Sigma}^{-1} (\mathbf{a} - \tilde{\mathbf{a}}) \right]\\
&= \prod_{x=1}^{k} \left( \frac{1}{\sqrt{2\pi} \tilde{\sigma}_x} \right) \exp \left( \frac{(a_x - \tilde{a}_x)^2}{\tilde{\sigma}_x^2} \right),
\end{align}
where $k$ is the number of fit parameters in model $M$, $\tilde{\Sigma}$ is the prior covariance matrix, and $\tilde{a}$ are the prior central values.
The second formula holds for the simplified case where the prior parameter covariance matrix $\tilde{\Sigma}$ is diagonal with entries $\tilde{\sigma}_x^2$.
Below will write $\chi_p^2$ to refer to the quantity in the exponent, $(\mathbf{a} - \tilde{\mathbf{a}})^T \tilde{\Sigma}^{-1} (\mathbf{a} - \tilde{\mathbf{a}})$.

The ordinary least-squares likelihood $\pr(D|\mathbf{a}, M)$ is normalized by the data covariance matrix and some factors of $(2\pi)$.
Since we are considering only the case of a fixed data set, this overall normalization is the same for all models and can be ignored here.
On the other hand, we retain for the present discussion the normalization of the prior distribution $\pr(\mathbf{a} | M)$, which differs for models with different numbers of parameters.

The best-fit point $\mathbf{a}^\star$ maximizes the above likelihood or, equivalently, minimizes the negative log-likelihood function
\beq \label{eq:chi2_aug}
-2 \log(\pr(D|\mathbf{a}, M) \pr(\mathbf{a} | M)) = \chi^2 + \chi_p^2 \equiv \chi_{{\rm aug}}^2.
\eeq
with the combination of terms defining the ``augmented chi-squared'' function \cite{Lepage:2001ym}.

\subsection{Bias correction of the model weights}\label{sec:bias}

Estimator bias occurs when a sample estimator differs from the ``true" underlying population value it is approximating.  A common example of such a bias occurs in the naive estimator of variance for data drawn from an underlying Gaussian distribution.
In this example, the bias is of order $1/N$, where $N$ is the sample size, which means that it will be automatically removed in the limit of large $N$.  

However, there are also examples of {asymptotic biases} which do not vanish as $N \rightarrow \infty$.
The sample maximum likelihood estimate (MLE) of the log likelihood function, $\chi_{\rm aug}^2(\mathbf{a}^\star)$, suffers from such an asymptotic bias.
Roughly speaking, because the MLE maximizes the sample log-likelihood, it tends to overshoot the true asymptotic value.
Fundamentally, this bias arises not from 
the choice of the MLE but rather from finite-sample-size fluctuations in the data itself.

To ensure convergence to the correct asymptotic model weight, we define the {bias-corrected} model weight to be \cite{Takeuchi:1976, Stone:1977, Shibata:1989, KonishiKitagawa:1996}
\beq
\pr(M|D)_{BC} = 
    \exp \left( -\tr[J^{-1}(\mathbf{a}^\star) I(\mathbf{a}^\star)] \right)
    \times
    \int d\mathbf{a} \frac{\pr(D|\mathbf{a},M) \pr(\mathbf{a}|M) \pr(M)}{\pr(D)}
    \label{eq:model_avg_BC}
\eeq
where $I$ and $J$ are the sample estimates of the log-likelihood Fisher information matrix and the (negative) Hessian matrix, respectively.
These matrices are defined in \cref{sec:bias-appendix}, which also gives an informal derivation of the bias correction term, $\sim\tr[J^{-1}(\mathbf{a}^\star) I(\mathbf{a}^\star)]$.  

When the model $M$ is correctly specified, i.e.~assuming that the ``true model'' from which the data are drawn can be described by $M$, it is straightforward to show (see \cite{Stone:1977,Shibata:1989} and \cref{sec:bias-appendix}) that as $N \rightarrow \infty$, the matrices $I$ and $J$ become identical, so that $J^{-1}(\mathbf{a}^\star) I(\mathbf{a}^\star) \rightarrow 1$, the $k\times k$ identity matrix.
For lattice gauge theory applications where the theoretical model rests on a firm physical foundation, the assumption of correct specification is likely to hold, and the bias correction simply becomes $\exp(-k)$, counting the total number of parameters in the model.

The appearance of the structure $\tr[J^{-1} I]$ is closely related to the Akaike information criterion \cite{akaike1998information,Akaike:1978} and its generalization, the Takeuchi information criterion \cite{Takeuchi:1976}. 
We discuss this connection more below.

\subsection{Gaussian approximation}

By construction, the sample likelihood $\pr(\mathbf{a}|D)$ is locally maximized at the best-fit parameter values $\mathbf{a}^\star$. Taylor expansion about the best-fit point gives
\beq
\chi_{{\rm aug}}^2(\mathbf{a})
\approx \chi_{{\rm aug}}^2(\mathbf{a}^\star)
+ (\mathbf{a} - \mathbf{a}^\star)^T {\Sigma^\star}^{-1} (\mathbf{a} - \mathbf{a}^\star)
+ ...
\eeq
where $\Sigma^\star$ is the standard best-fit covariance matrix evaluated at the best-fit point,
\beq
{\Sigma_{xy}^\star}^{-1} \equiv \frac{1}{2} \left. \frac{\partial^2 \chi_{{\rm aug}}^2}{\partial a_x \partial a_y}\right|_{\mathbf{a} = \mathbf{a}^\star}
\eeq
This approximation, known in the probability literature as the ``Laplace approximation," becomes increasingly accurate in the limit of large statistics in the data set $D$. 
(For linear models, as defined in \cref{sec:linear_vs_nonlinear}, this approximation is of course exact, since the $\chi^2$ function is quadratic in parameters $\mathbf{a}$.)
Within this approximation, the integral for model weight becomes Gaussian and can be evaluated analytically:
\begin{align}
\pr(M|D) &= \int d\mathbf{a} \frac{\pr(D|\mathbf{a},M) \pr(\mathbf{a}|M) \pr(M)}{\pr(D)} \\
&\approx \int d\mathbf{a}\ \frac{1}{(2\pi)^{k/2} (\det \tilde{\Sigma})^{1/2}} e^{-{\chi_{{\rm aug}}^2}(\mathbf{a}^\star)/2 - \frac{1}{2} (\mathbf{a} - \mathbf{a}^\star)^T {\Sigma^\star}^{-1} (\mathbf{a} - \mathbf{a}^\star)} \frac{\pr(M)}{\pr(D)} \\
&= \frac{\pr(M)}{\pr(D)} (2\pi)^{-k/2} (\det \tilde{\Sigma})^{-1/2} \left[ (2\pi)^{k/2} e^{-{\chi_{{\rm aug}}^2}(\mathbf{a}^\star)/2} (\det {{\Sigma^\star}^{-1}})^{-1/2} \right] \\
&= \frac{\pr(M)}{\pr(D)} (\det \tilde{\Sigma})^{-1/2} e^{-{\chi_{{\rm aug}}^2}(\mathbf{a}^\star)/2} (\det {\Sigma^\star})^{1/2}  \label{eq:approxPMD}.
\end{align}
Thus, neglecting the term $\pr(D)$ which is the same for all models, one finds the following form for the log-likelihood
\beq
-2 \log(\pr(M|D)) \approx -2 \log(\pr(M)) + {\chi_{{\rm aug}}^2}(\mathbf{a}^\star) + \log \det \tilde{\Sigma} - \log \det {\Sigma^\star},
\eeq
Including the bias correction term introduced in \cref{eq:model_avg_BC} gives the overall result
\begin{align}
&-2 \log(\pr(M|D)_{\rm BC}) = -2 \log(\pr(M)) + {\chi_{{\rm aug}}^2}(\mathbf{a}^\star) + \log  \frac{\det \tilde{\Sigma}}{\det {\Sigma^\star}} + 2\ \tr[ J^{-1}(\mathbf{a}^\star) I(\mathbf{a}^\star) ]. \label{eq:GAPfull}
\end{align}

From \cref{eq:GAPfull}, we see that the posterior probability $\pr(M|D)$ encapsulates the principle of Occam's Razor: models with large $\chi_{{\rm aug}}^2(\mathbf{a}^\star)$ are penalized, but so are models which have a large number of free parameters.
(As discussed above, the final bias-correction term reduces to $2k$ asymptotically, where $k$ is the number of parameters.)

Unfortunately, in the presence of any empirical priors and/or models with differing dimensions, an effect known as the Jeffreys-Lindley paradox \cite{jeffreys1998theory, Cousins:2013hry, Lindley:1957aa} leads to out-sized dependence on the prior widths that can severely distort the overall results.  The presence of the paradox is linked closely to the use of ``empirical priors'' which are not based on true prior information and can be taken to be arbitrarily wide.  Note that for estimation of the best-fit parameters $\mathbf{a}^\star$ in a fixed model there is no such distortion, and the use of empirical priors is not problematic.
However, when the normalization of the likelihood is important the paradox can lead to nonsensical results if the prior width is taken to be extremely large.

For the particular case at hand, instead of attempting to confront the Jeffreys-Lindley paradox head-on, we will instead argue that the effect of the $(\log \det \tilde{\Sigma} - \log \det \Sigma^\star)$ terms will vanish asymptotically and so can be viewed as a $1/N$ effect in the sample size $N$.
For the sake of argument, imagine adopting the following cross-validation procedure:
partition our full data set $D$ into a small ``training set'' $D_T$, and the remaining data $D_T^c$. 
Then imagine first fitting the training set $D_T$ to determine a set of fit parameters $\mathbf{a}_T$. 
These results for $\mathbf{a}_T$ can then be used to fix priors on $\mathbf{a}$ for the fit to the remaining data $D_T^c$.  
In practice, the cost of such a procedure is that it would ``use up'' the data in $D_T$ to fix priors, reducing the overall statistical precision of the analysis.
However, in the asymptotic limit $N \rightarrow \infty$, both partitions $\{D_T, D_T^c\}$ approach the true asymptotic distribution and will yield exactly the same fit results.
Therefore, in this limit $\tilde{\Sigma} \rightarrow \Sigma^\star$ and the determinant terms in \cref{eq:GAPfull} cancel completely.

In other words, this argument establishes that the effect of the determinant terms is subleading in the asymptotic limit, and they may be dropped even without explicit use of a cross-validation procedure.
In the same $N\to\infty$ limit, the bias-correction term reduces to twice the number of parameters (see \cref{sec:bias-appendix}), and we recover the well-known Akaike information criterion (AIC) \cite{akaike1998information,Akaike:1978}:
\beq
-2 \log(\pr(M|D)_{\rm BC}) \xrightarrow{N \rightarrow \infty} \rm{AIC}_M = -2\log (\pr(M)) + \chi_{{\rm aug}}^2(\mathbf{a}^\star) + 2k.\label{eq:AIC_M}
\eeq
This is our main result for the bias-corrected model weights in the limit $N \rightarrow \infty$.  We advocate use of \cref{eq:AIC_M} in all cases; without a more complete treatment of finite-sample-size effects, there is no guarantee that inclusion of the determinant terms in \cref{eq:GAPfull} will improve estimation of the model weights.
From this point forward, we adopt the bias-corrected form as the default choice of $\pr(M|D)$ unless explicitly stated otherwise and thus drop the ``BC'' subscript.

More generally, simply dropping the determinant terms from~\cref{eq:GAPfull} gives the Takeuchi information criterion (TIC)~\cite{Takeuchi:1976}.
However, as discussed in \cref{sec:bias-appendix}, the TIC form of the bias correction is only necessary in the case of model mis-specification.  For models that obviously fail to describe the data, the $\chi_{{\rm aug}}^2$ term in either information criterion will dwarf the size of the bias correction, so the distinction between TIC and AIC is most important in cases where little information is available about the true model.
In the present context, strong physical motivation often exists for trusting the correctness of the models in use.
We therefore advocate the use of the AIC for general model averaging purposes in lattice gauge theory.

\subsection{Practical example: polynomial data}
\label{sec:ex1}

\begin{figure}
\centering
\includegraphics[width=0.65\textwidth]{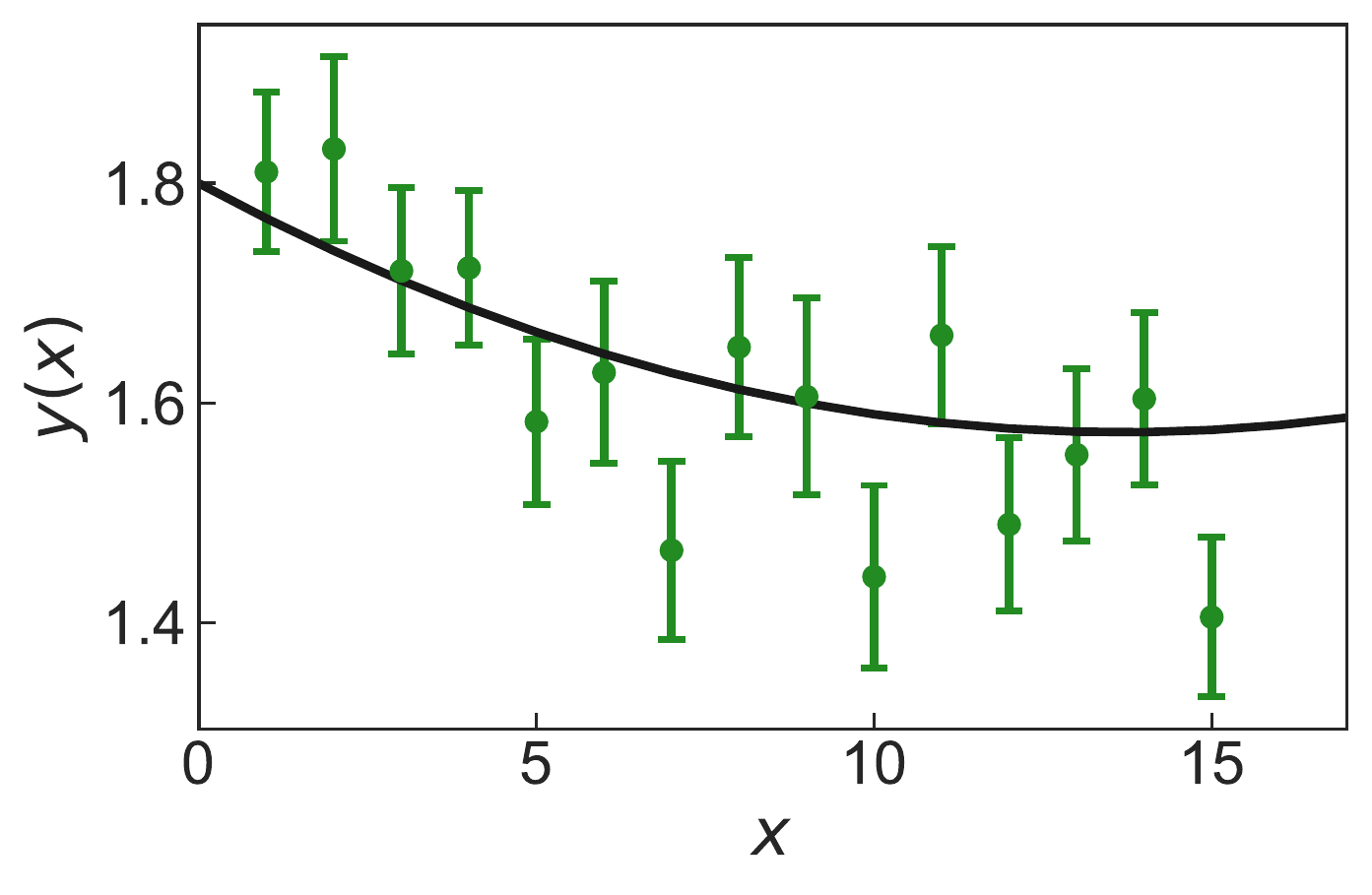}
\includegraphics[width=0.9\textwidth]{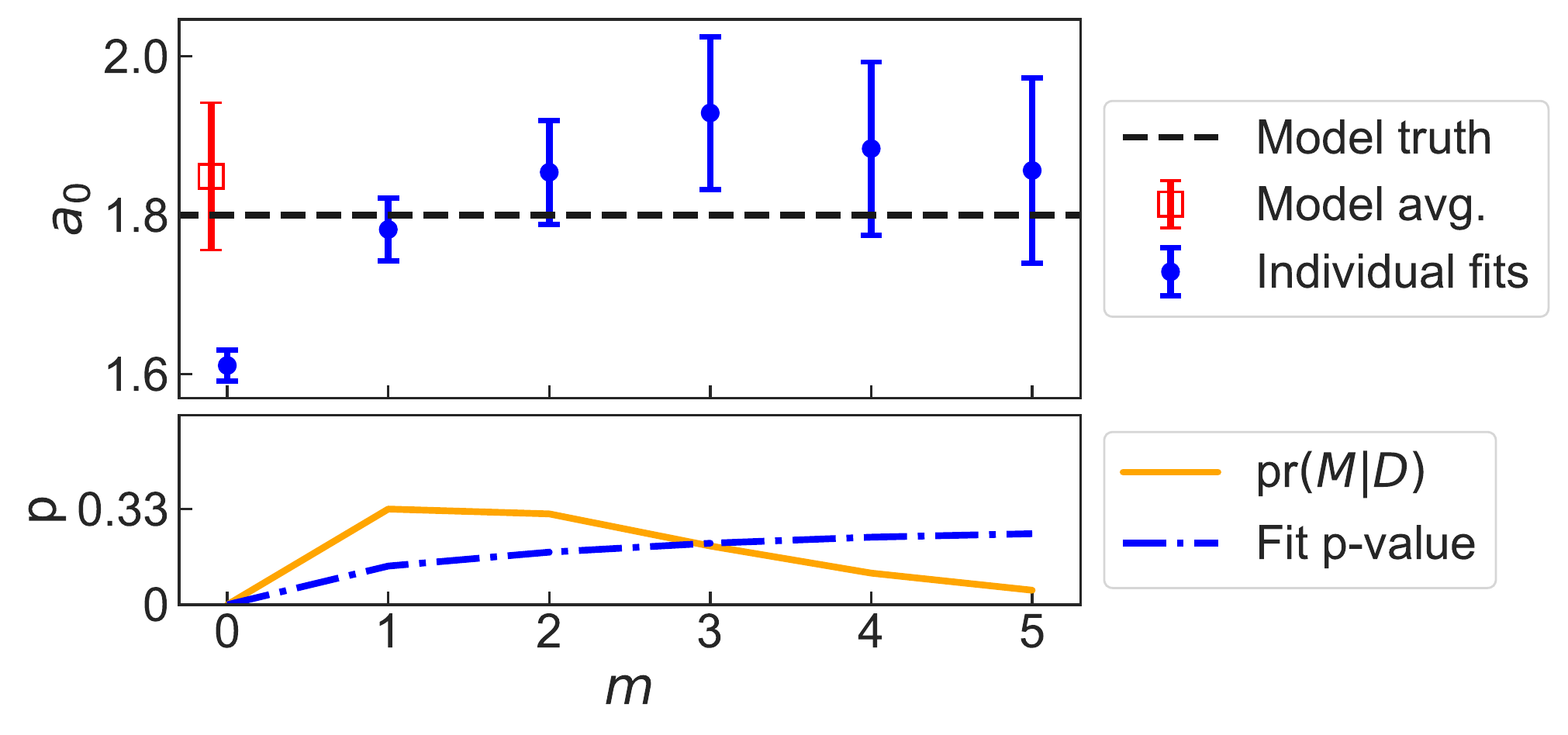}
\caption{Top: synthetic data (green points) for the given quadratic ``model truth'' function (black curve) plus Gaussian noise, with noise sample size $N = 160$. Bottom: fit results for degree-$m$ polynomial models (blue circles), compared to the known value $a_0 = 1.8$ (black dashed line).  The model-averaged result (red open square) obtained from the weighted average of the blue fits is shown at $m=0$.  The lower inset shows the standard $p$-value (blue dashed line) and the model weight calculated from the AIC (orange solid line). Comparison of these curves shows the ``Occam's razor'' effect, with the AIC penalizing fits with roughly equal goodness of fit but more fit parameters.\label{fig:poly_test}}
\end{figure}

To demonstrate the method and some key features, we begin by considering a simple toy model.  We begin by specifying a quadratic ``model truth" polynomial function,
\beq
F(x) = 1.80 - 0.53 \left( \frac{x}{16} \right) + 0.31 \left( \frac{x}{16} \right)^2.
\eeq
A set of $N$ mock data samples are generated for $x \in \{1,2,...,16\}$ by taking the model truth for each point and adding Gaussian noise $\eta(x)$, uncorrelated in $x$ with mean zero and standard deviation $\sigma = 1.0$.
The resulting mock data $y(x) = F(x) + \eta(x)$ are plotted in \cref{fig:poly_test} (top panel) for the choice $N=160$; we will also study the $N$-dependence.  \footnote{The code used to generate this practical example, as well as the synthetic correlation function example below, is publicly available at \url{https://github.com/etneil/model_average_paper/}.}

We take as our space of model functions polynomials labeled by their degree $m$,
\beq
f_m(x) = \sum_{j=0}^m a_j \left( \frac{x}{16} \right)^j.
\eeq
with $0 \leq m \leq 5$.  We take the flat prior $\pr(m) = 1/6$, corresponding to minimal prior knowledge about the functional form of the model (except that it is polynomial.)  All parameter priors are taken to be Gaussian with mean 0 and width 10 (the parameters are, essentially, unconstrained.)

\begin{table}
\begin{tabular}{|c|cccccc|}
\hline
&$m=0$&$m=1$&$m=2$&$m=3$&$m=4$&$m=5$\\
\hline
 $a_0$    & 1.640(20) & 1.782(39)  & 1.854(65) & 1.929(96) & 1.88(11)  & 1.86(12)  \\
 $a_1$    &           & -0.339(68) & -0.74(30) & -1.56(83) & -0.8(1.2) & -0.5(1.3)  \\
 $a_2$    &           &            & 0.39(28)  & 2.4(1.9)  & -1.0(4.4)  & -1.7(4.5)  \\
 $a_3$    &           &            &           & -1.3(1.3) & 4.2(6.4)  & 2.7(6.8)   \\
 $a_4$    &           &            &           &           & -2.8(3.3) & 1.6(7.4)   \\
 $a_5$    &           &            &           &           &           & -2.7(4.0)  \\
 \hline
 $\chi_{{\rm aug}}^2$  & 41.25     & 16.11     & 14.22     & 13.09      & 12.33     & 11.88      \\
 $p$-value & 0.00      & 0.37       & 0.51      & 0.60      & 0.65      & 0.69       \\
 AIC${}_m$  & 43.24     & 20.12      & 20.22     & 21.08      & 22.32     & 23.88      \\
 pr$(M|D)$ & 0.00      & 0.33       & 0.31      & 0.20      & 0.11       & 0.05       \\
\hline
\end{tabular}
\caption{Individual best-fit results and associated quantities for $N=160$.  The model-averaged value for the intercept is $\ev{a_0} = 1.849(93)$. \label{tab:poly_test}}
\end{table}

The results of the fits to individual models as well as the model-averaging results are shown in \cref{fig:poly_test} (bottom panel) and in \cref{tab:poly_test}.  Note that although all models with $m > 2$ provide a good description of the data in terms of $\chi^2$, the bias-corrected model probability estimated through the AIC places relatively more weight on the simpler models, with the maximum probability assigned to the correct choice $m=2$.  The model-averaged result is consistent with model truth and has slightly larger uncertainty than the individual fit to the correct model $m=2$.

\begin{figure}
\includegraphics[width=0.9\textwidth]{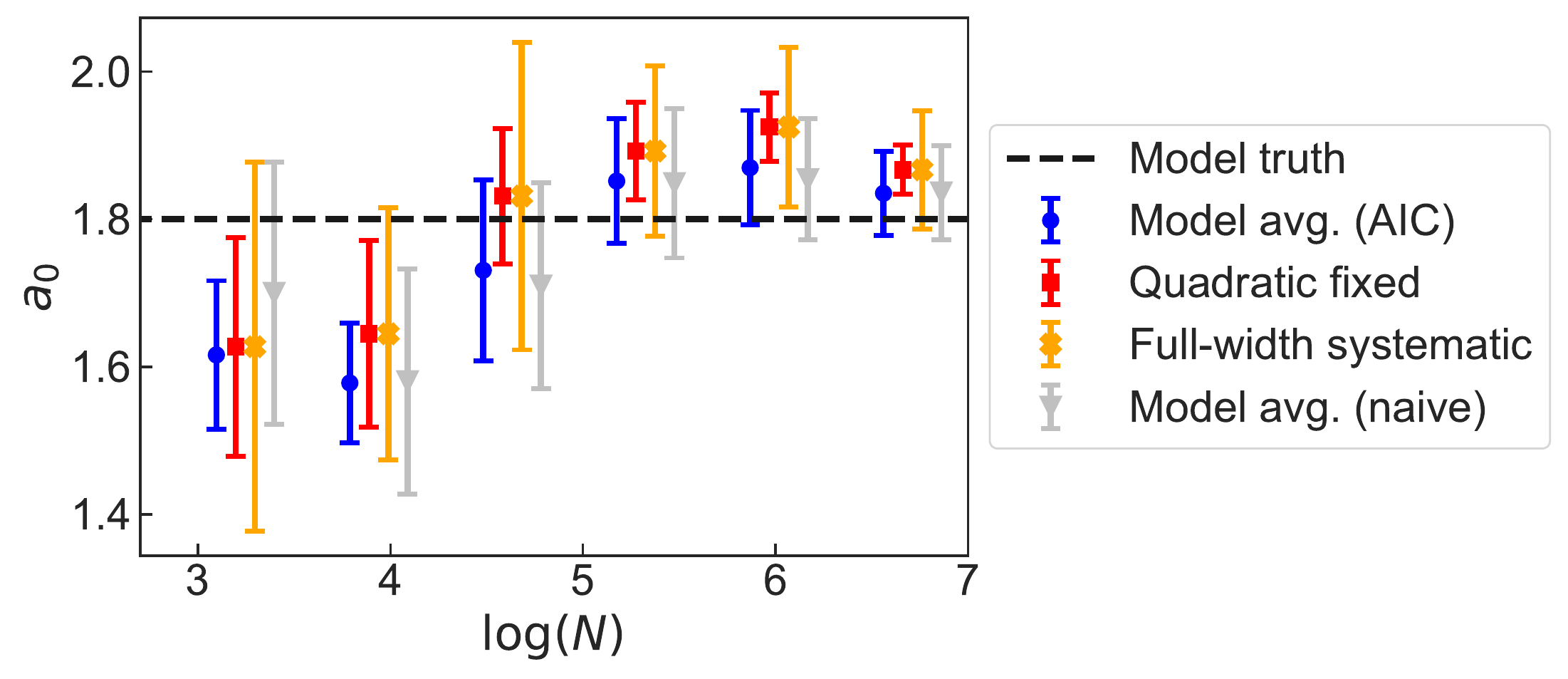}
\caption{Scaling of various estimates of the intercept $a_0$ vs. the data sample size $N$.  The true value (dashed line) is $a_0 = 1.8$. The blue circles (model average using the AIC) show good consistency with both the model truth and with the estimates using the correct quadratic model form (red squares).  Using the full-width difference between all models with fit $p$-value greater than $0.1$ as a systematic error (orange crosses) tends to give larger uncertainty than the AIC model average.  Finally, averaging using a ``naive'' estimate of $\pr(M|D)$ which omits the bias-correction term (silver triangles) does not directly lead to bias in the estimation of $a_0$, but also gives slightly larger uncertainty due to overweighting of more complicated models as discussed in the text.\label{fig:poly_scale}}
\end{figure}

As noted, the results so far use a fixed sample size of $N=160$.  We repeat the test as described above with several values of $N \in [20, 640]$, showing the final estimated result for $a_0$ using various procedures in \cref{fig:poly_scale}.  The result of the model averaging procedure using the AIC is seen to be consistent with model truth in all cases, with an error that is uniformly smaller than the more conservative procedure of taking the full variation of the mean over all models with $p > 0.1$ as a systematic error.  The AIC model-averaged error is larger than the error on the ``quadratic fixed'' result using the known true quadratic model; this is to be expected, as using model averaging rather than a fixed model necessarily builds in an additional systematic error due to model uncertainty (see the discussion in Sec.~\ref{sec:ma_formulas}.)  Although in this simple example the true model is known exactly, we emphasize that this situation is rare, and in the absence of such exact knowledge the use of fixed-model fits can result in underestimation of parameter errors.

Omitting the bias-correction term and averaging using only the $\chi^2$ results to estimate model probability (the ``naive'' estimate) also tends to give slightly larger error than model averaging using AIC, but the results remain consistent with the correct answer.  In the absence of the bias correction, the likelihood of the models with $m>2$ is overestimated.  However, because those higher-order polynomial models include the $m=2$ model within their parameter space, they tend to estimate the correct value of the intercept $a_0$ on average.  As a result, there is no bias introduced into the \emph{mean} result for $a_0$ when using biased model probabilities in this particular example.  However, the error bar is slightly overestimated when using the naive estimator.

\section{Data subset selection as a model variation problem}
\label{sec:subset}

A routine part of modeling lattice Monte Carlo data is data subset selection, i.e., choosing a ``cut'' on the data beyond which the model is not applied.
A canonical and simple example is fitting a two-point correlation function $C(t)$ to extract the ground-state energy.
The full model expected to describe this correlation function involves an infinite tower of exponentials, $\sum_i^\infty A_i e^{-E_i t}$.
In practice, one truncates the sum after a finite number of terms and then selects a minimum value $t_{\rm min}$ below which the data are simply ignored.
Choosing the precise value of $t_{\rm min}$ is generally done by hand. 

Although this process is typically thought of as data selection problem, 
it can easily be reformulated as a \emph{model} selection problem.
In the previous example, the justification for ignoring data below $t_{\rm min}$ is partially one of expediency.
If we are only interested in the first few states, it suffices to look at times with $t \ge t_{\rm min}$ where they dominate.
Times below $t_{\rm min}$ will be heavily contaminated by contributions from the higher excited states, and little to no information about the first few states is lost by ignoring them.

Based on this observation, we can define a joint model that describes the full data set.
First, select a subset of the data and imagine fitting the model of choice $M$ to this subset as usual.
Second, imagine fitting the remaining data to a ``perfect" model with zero degrees of freedom.
For example, the ``perfect" model could be a polynomial with degree equal to $N_{\rm cut}$, but in principle other functional forms will also work.
Because the ``perfect'' model has zero degrees of freedom, there exists a solution for its parameters for which the differences between the model and the sample means vanish exactly.

To give an explicit construction, we first define a partition $P$ of the data vectors into 
$y_i = (y_i^{\rm cut}, y_i^{\rm keep})$, 
where
$y_i^{\rm keep}$
are the subset to be modeled and
$y_i^{\rm cut}$
are the cut data.
We then define the corresponding partitioned model $g_M(\mathbf{a},P)$ as
\beq
y_i - g_M(\mathbf{a},P) = 
    \begin{cases}
        y_i - \bar{y}^{\rm cut} ,&\ y_i \in y_i^{\rm cut} \\
        y_i - f_M(\mathbf{a}), &\ y_i \in y_i^{\rm keep},
    \end{cases}
\eeq
where 
$\bar{y}^{\rm cut}$
is the sample average of the
$y_i^{\rm cut}$.
The partition-dependent log likelihood is then, dropping constant terms that do not change with fixed data set $D$,
\begin{align}
-2 \log \pr(D|\mathbf{a}, M) = \sum_{i=1}^N \chi_i^2(P)
    &= \sum_{i=1}^N (y_i - g_M(\mathbf{a},P))^T \Sigma^{-1} (y_i - g_M(\mathbf{a},P)) \\
    &= \sum_{i=1}^N (y_i^{\rm keep} - f_M(\mathbf{a}))^T \Sigma_P^{-1} (y_i^{\rm keep} - f_M(\mathbf{a})) +  (\rm{const})
\end{align}
where $\Sigma_P^{-1}$ is the submatrix of the full inverse data covariance matrix $\Sigma^{-1}$ which corresponds to the data subset 
$y_i^{\rm keep}$.
All other terms involving the cut data contain the expression
$\bar{y}^{\rm cut} - g_M(\mathbf{a}, P)$ 
at least once and therefore vanish identically by construction, even in the presence of off-diagonal correlations between
$y_i^{\rm keep}$
and
$y_i^{\rm cut}$.

Since matrix inversion does not generally commute with subspace projection, the matrix $\Sigma_P^{-1}$ typically differs from $(\Sigma_P)^{-1}$, the inverse of the covariance sub-matrix.
However, in practical lattice applications $(\Sigma_P)^{-1}$ is often used as an approximation to $\Sigma_P^{-1}$; the difference between these matrices is given by terms that are suppressed 
whenever long-range (i.e., further off-diagonal) correlations are generally smaller than short-range ones.
An obstruction to using $\Sigma_P^{-1}$ directly is that finite-sample estimates of the full covariance matrix $\Sigma^{-1}$ are typically ill-conditioned.
Therefore, in what follows we will use $(\Sigma_P)^{-1}$.

The result of this construction is that the contribution from the ``perfect'' model describing the data outside the chosen subset is $\Delta \chi_{\rm aug}^2 = 0$.  However, there remains a bias-correction term which accounts for the $N_{\rm cut}$ additional model parameters used to describe the cut data 
$y_i^{\rm cut}$.  
The bias correction is still necessary because although the ``perfect'' model exactly describes the data sample as given, the values of the perfect model parameters will fluctuate as additional data is added.
The difference between the perfect model parameters at finite sample size and their asymptotic values leads to a bias correction as described in \cref{sec:bias-appendix}.
We emphasize that the bias correction term itself does not vanish since the individual terms $\chi_i^2(\mathbf{a})$ do not vanish identically for the perfect model, only the sum.
Thus, the overall model probability for the joint model is easily seen to be obtained from the modified expression
\beq\label{eq:model_avg_cut}
\rm{AIC}_{M, N_{\rm cut}} = -2\log (\pr(M)) + \chi_{{\rm aug}}^2(\mathbf{a}^\star) + 2k + 2N_{\rm cut},
\eeq
where $\chi_{{\rm aug}}^2(\mathbf{a}^\star)$ is evaluated only for the model $M$ and for data within our selected subset.  

The result \cref{eq:model_avg_cut} depends only on quantities that may be estimated from the subset model fit, and on counting factors.  As a result, in practice we do not need to construct the ``perfect'' model at all.  We  note that a similar penalty term for removal of data points was also proposed in \cite{Borsanyi:2020mff} in a frequentist context.

Although we were motivated by the example of a two-point correlator where the data are cleanly divided into two subsets along a single dimension, the argument above holds for arbitrarily complex subdivisions of the full data set.  Whatever subset of the data we choose to fit explicitly to model $M$, the joint model which also describes the remaining data will contribute an additional factor of $2N_{\rm cut}$ to the information criterion.  We can also consider a set of models $\{M\}$ and perform ordinary model averaging over the joint space defined by $\{M\}$ and the parameters that uniquely define a data subset.

\subsection{Practical example: Synthetic correlation functions}
\label{sec:ex2}

To test the data subset selection procedure, we set up another toy-model example resembling a two-point correlation function, following the example description above.  The ``model truth'' in this case is a two-state exponential,
\beq
F(t) = 2.0 e^{-0.8 t} + 10.4 e^{-1.16t}.
\eeq
To generate synthetic data, we add correlated Gaussian noise $\eta(t)$ with mean zero and variance 0.09.  The noise is added fractionally to the data, i.e., the synthetic data are generated according to the formula $y(t) = F(t) (1 + \eta(t))$.
The correlation matrix of the noise takes the form $\rho_{t,t'} = \rho^{|t-t'|}$, i.e. equal to 1 on the diagonal and decreasing according to a power law as the temporal separation between points increases, similar to a real lattice QCD correlation function.
We fix the numerical correlation coefficient $\rho = 0.6$. $N$ mock data samples are generated for $t \in \{0, 1, ..., 31\}$.

Additional trials in which the above parameters have been varied were also tried, including using uncorrelated Gaussian noise instead of correlated.  No qualitative difference in the outcome of the tests was observed with these variations.

For this test, we consider a single model which consists of a single exponential term,
\beq
f(t) = A_0 e^{-E_0 t}.
\eeq
This model is fit to all data in the range $[t_{\rm min}, 31]$.  We consider all values of $t_{\rm min}$ from 1 to 28, with the goal of using model averaging with \cref{eq:model_avg_cut} to obtain a combined result for the ground-state energy $E_0$. 

\begin{figure}
\begin{subfigure}{}
\includegraphics[width=0.64\textwidth]{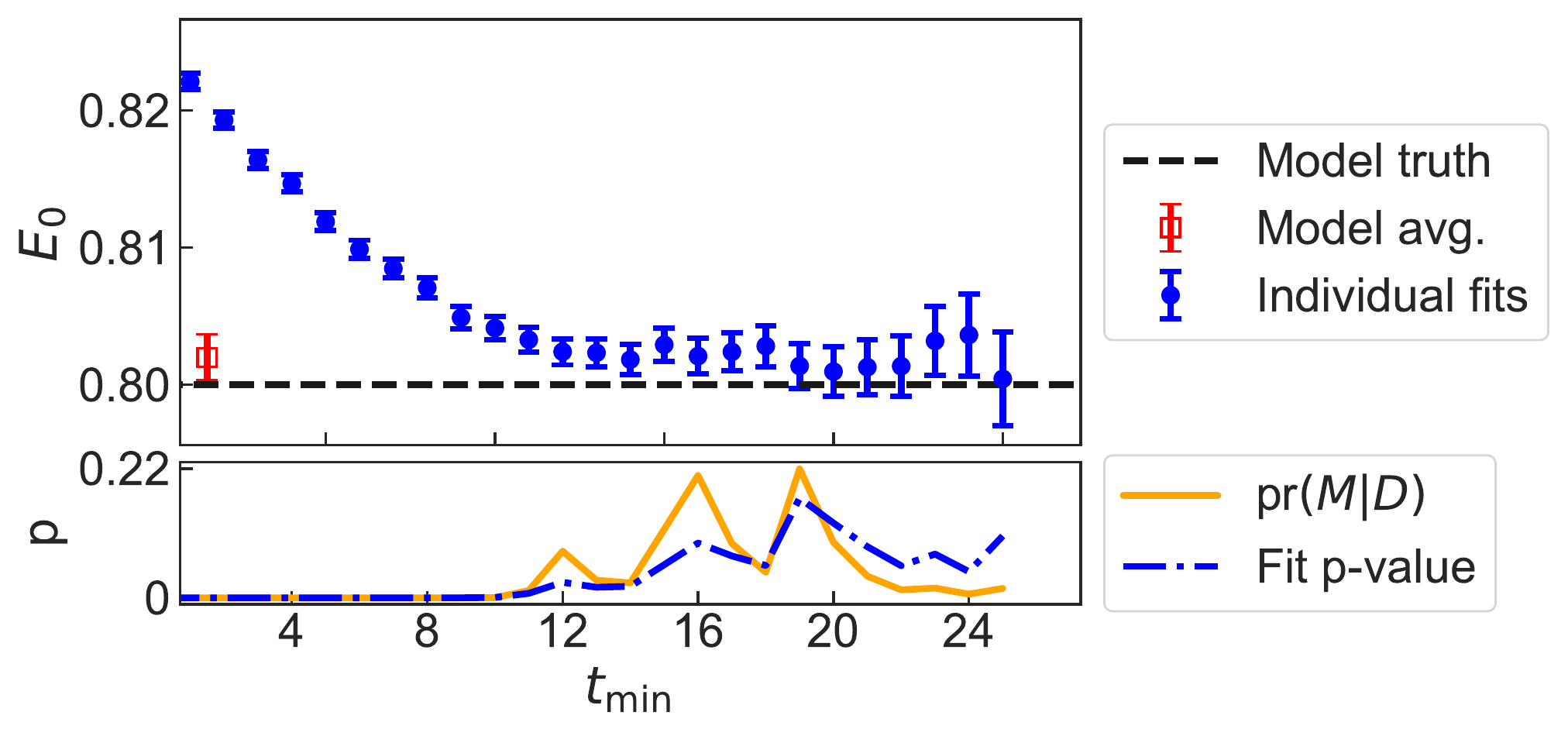}
\includegraphics[width=0.64\textwidth]{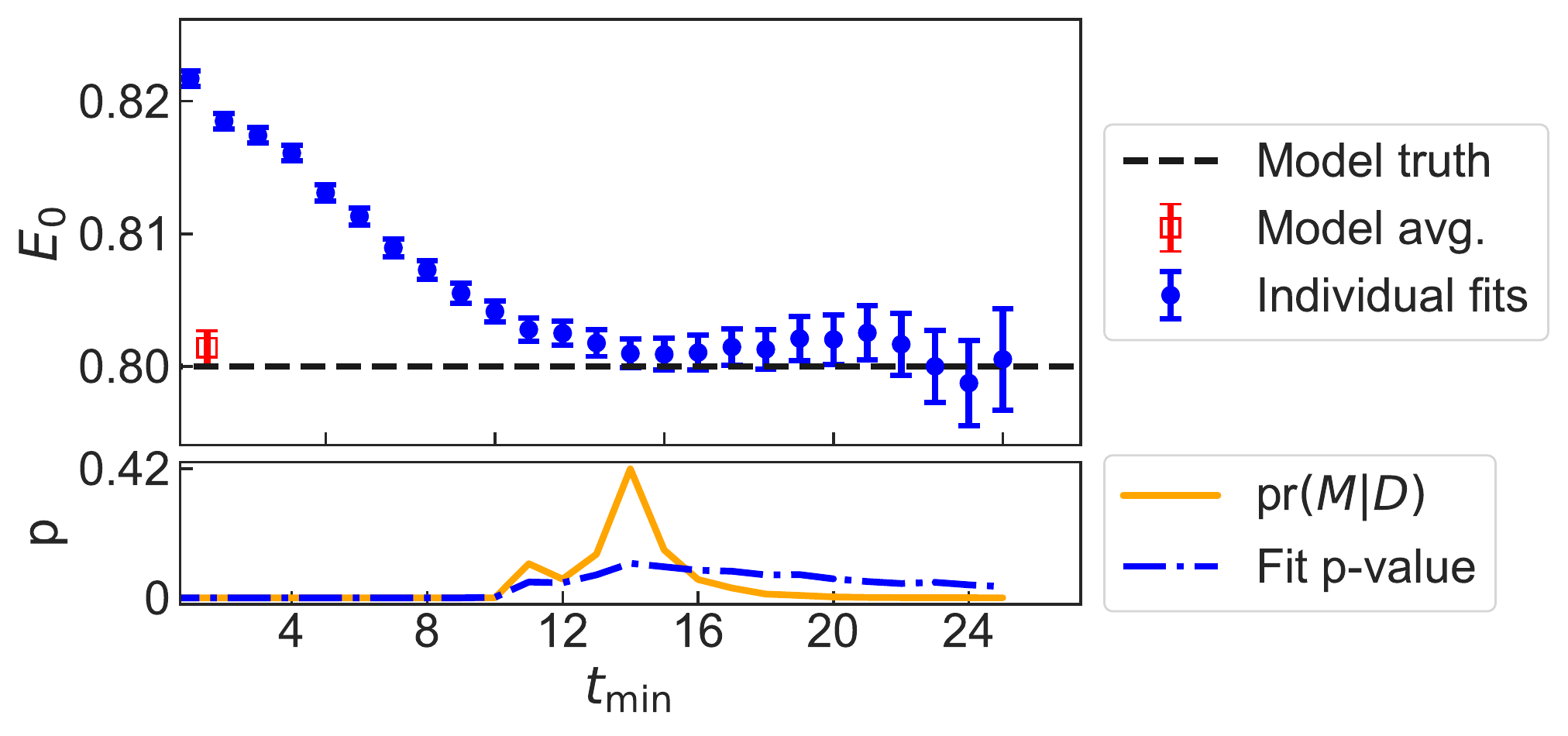}
\end{subfigure}
\begin{subfigure}{}
\includegraphics[width=0.64\textwidth]{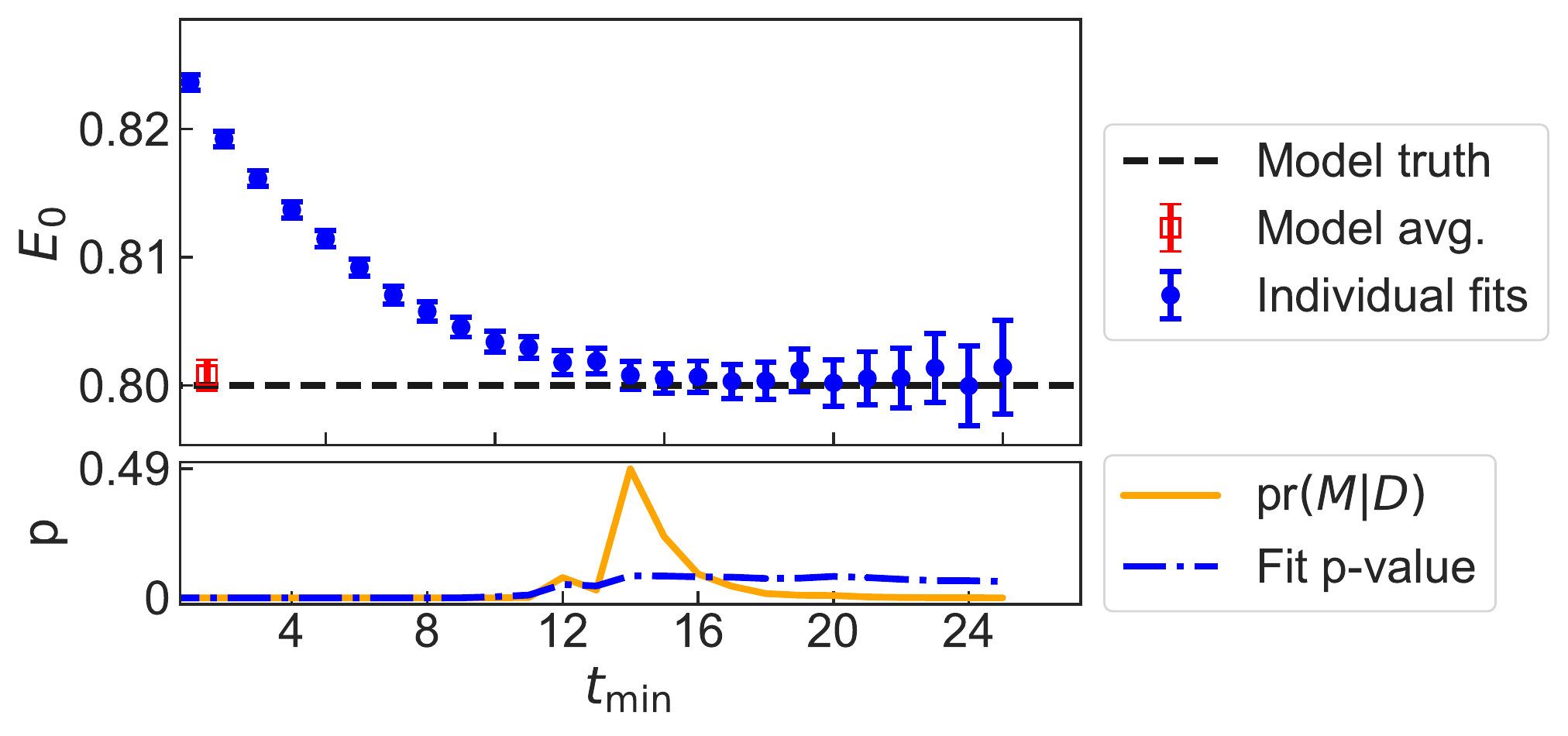}
\includegraphics[width=0.64\textwidth]{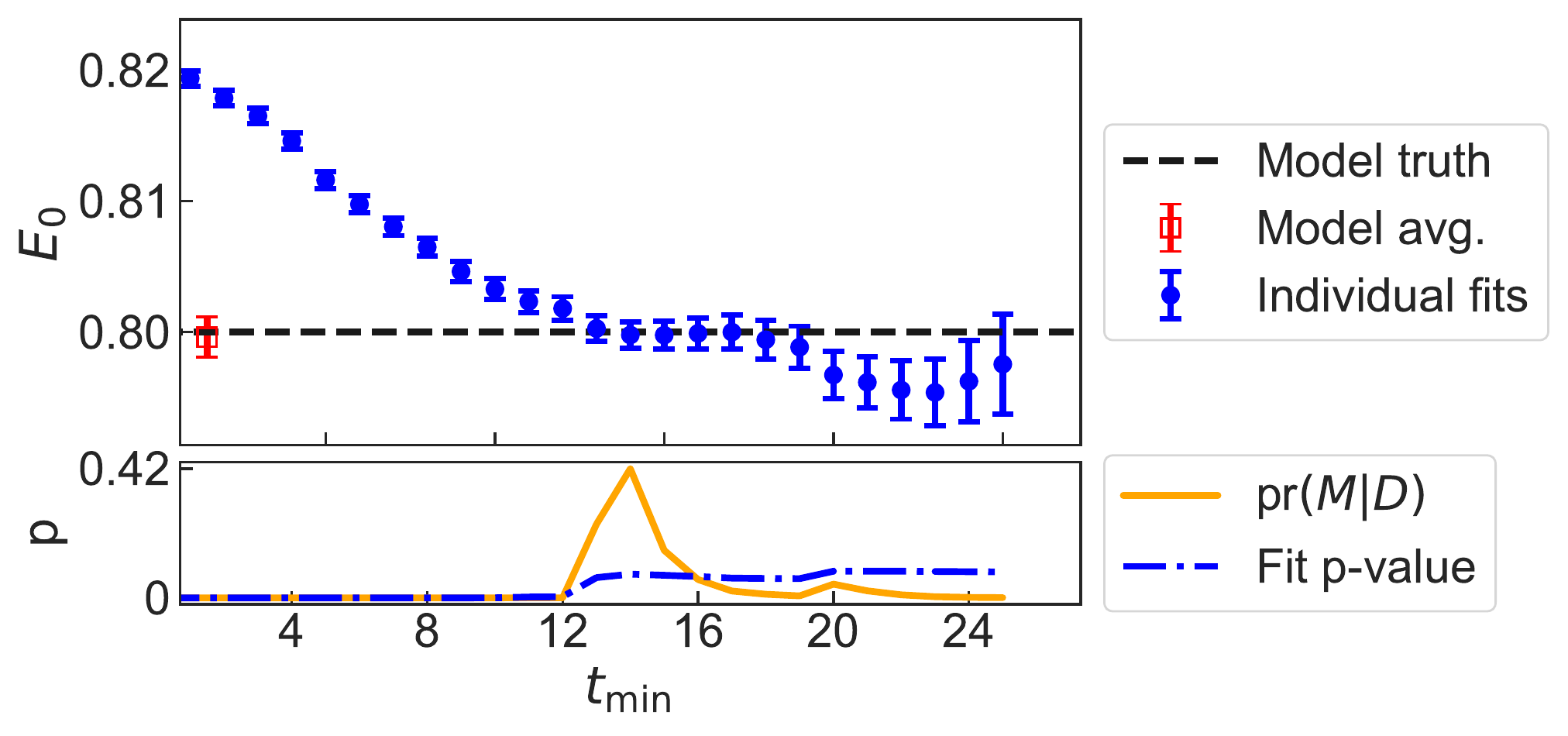}
\end{subfigure}
\caption{Fit results for the ground-state energy with true value $E_0 = 0.8$ (black dashed line), with the data cut away below $t_{\rm min}$ (blue points).  The model-averaged result (red open square) shows good agreement with model truth in all cases.  The lower inset shows the standard $p$-value (blue dashed line) and the model weight calculated from \cref{eq:model_avg_cut} (orange solid line).  The four subfigures represent four random draws of correlated Gaussian noise, but are otherwise identical.
\label{fig:exp_test}}
\end{figure}

The results of four independent trials following the above procedure with $N=500$ are shown in \cref{fig:exp_test}.  Excited-state contamination, i.e. the influence of the second exponential state which is not present in our fit model, is clearly visible at low $t_{\rm min}$.  In each case, excellent agreement of the model-averaged result with model truth is seen.  As in the polynomial example, the bias-corrected model probability is seen to weight simpler models more strongly, which in this case means favoring fits that cut away less of the data.

\begin{figure}
\includegraphics[width=0.9\textwidth]{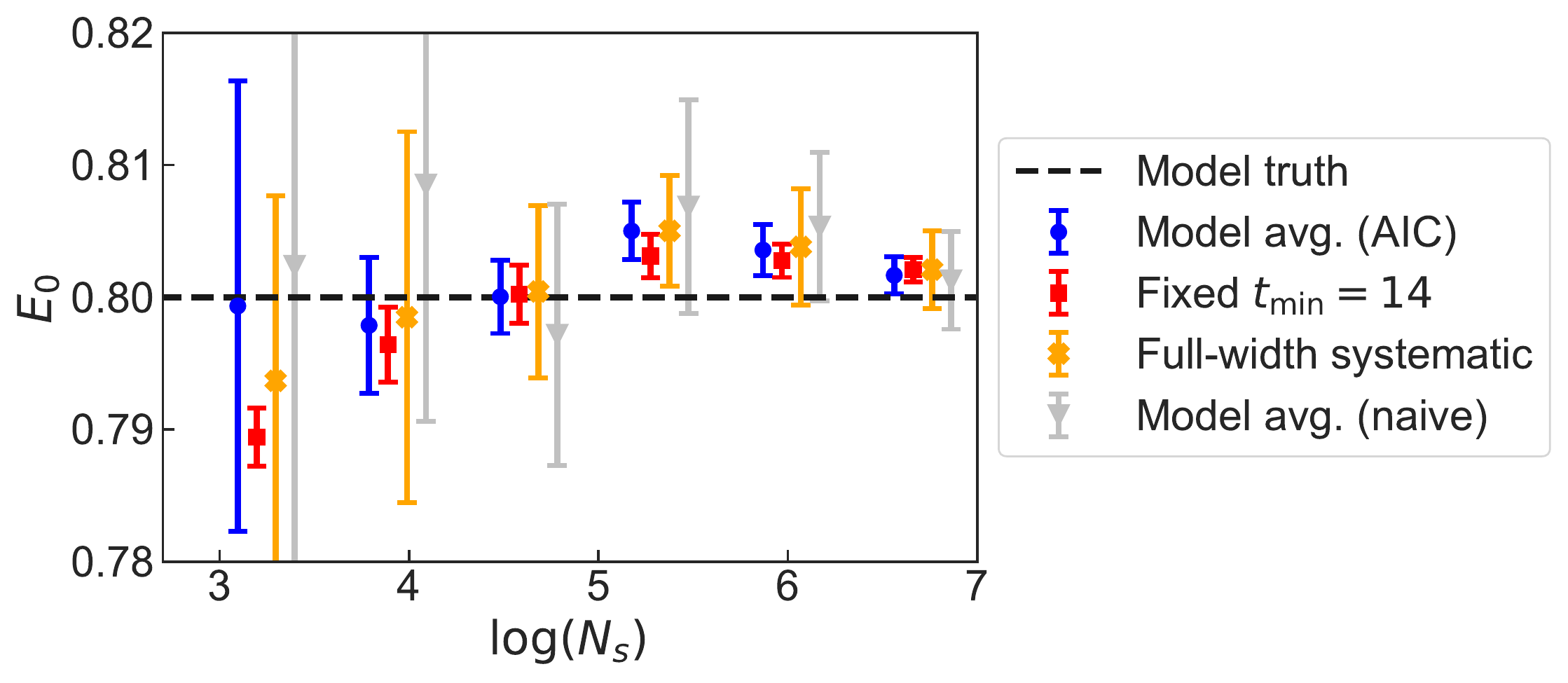}
\caption{Scaling of various estimates of the ground-state energy $E_0$ vs. the data sample size $N$.  The true value (dashed line) is $E_0 = 0.8$. The blue circles (model average using the AIC) show good consistency with the model truth and generally comparable error to fitting with fixed $t_{\rm min}$ (red squares).  Using the full-width difference between all models with fit $p$-value greater than $0.1$ as a systematic error (orange crosses) tends to give significantly larger uncertainty than the AIC model average.  Finally, averaging using a ``naive'' estimate of $\pr(M|D)$ which omits the bias-correction term (silver triangles) also leads to significantly larger uncertainty due to overweighting of more complicated models as discussed in the text. \label{fig:exp_scale}}
\end{figure}

In \cref{fig:exp_scale}, we repeat the above exercise while varying the sample size $N$, once again over the range $N \in [20, 640]$.  The results of model averaging using the AIC, i.e. using \cref{eq:model_avg_cut}, show good consistency with the known result.  Although the error of the model-averaged result is generally somewhat larger than the error for using a single fixed choice of $t_{\rm min}$, the latter has an unaccounted-for systematic error due to model truncation.  (Indeed, if we do not adjust $t_{\rm min}$ as $N \rightarrow \infty$, we expect the result for $E_0$ to become incompatible with the correct ground-state energy, as the contamination from the second state will eventually be resolved in a large enough sample.)

On the other hand, the error on the model-averaged result is generally much smaller than the more conservative full-width estimate, resulting from taking the full variation of mean results over all models with $p > 0.1$ as a systematic error as in the polynomial example \cref{sec:ex1}. In contrast to the polynomial example, here omission of the important bias-correction term in the AIC (i.e. dropping the $N_{\rm cut}$ contribution to model weight) causes a drastic inflation of the error in the averaged result for $E_0$.  Once again, the interpretation is that omitting the bias correction causes overestimation of the likelihood for the more complicated models, in this case models with larger $t_{\rm min}$.  These models generally give results for $E_0$ consistent with the correct answer, but with much larger errors, and the model-averaged result is altered accordingly.

\subsection{Practical example: QCD correlation functions}
\label{sec:ex3}
\subsubsection{Masses from a two-point correlation function}

We now consider the example of model averaging applied to a pion two-point correlation function from a real lattice QCD calculation.
This correlator has been used in published work by the Fermilab Lattice and MILC collaborations~\cite{Bazavov:2018kjg}.
The underlying gauge-field ensemble used has a lattice spacing of $a \approx 0.09$ fm and a pion mass of about 215 MeV.
In this example, the correlation function was constructed using staggered fermions and corresponds to a pion with energy $E_\pi \approx 300$ MeV.

\begin{figure}[th]
\centering
\includegraphics[height=0.85\textheight]{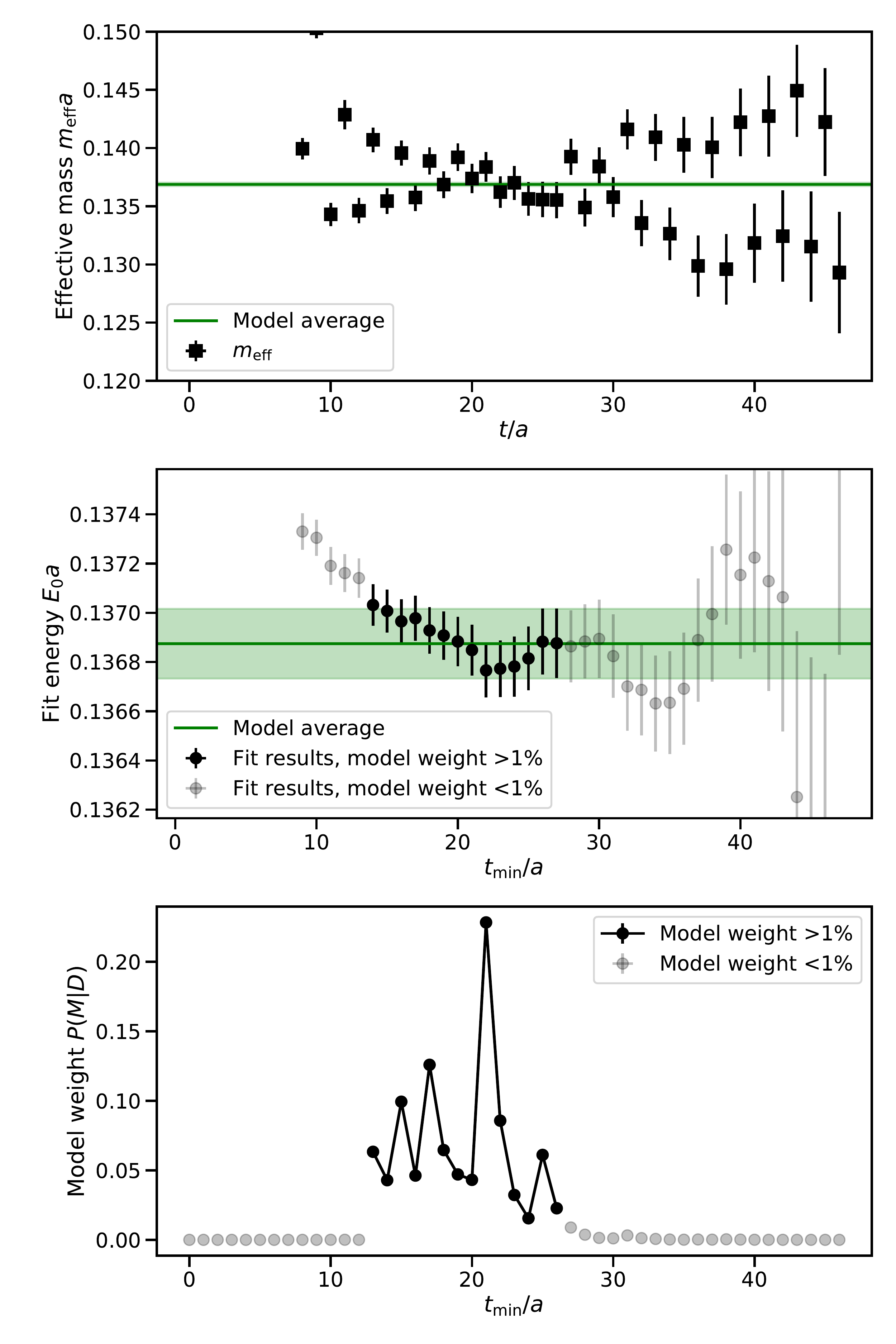}
\caption{
Model averaging results for a pion two-point correlation function using staggered fermions.
\textbf{Top:} The effective mass in lattice units and the final result of model averaging.
The oscillating contributions are from the opposite-parity states associated with staggered fermions.
\textbf{Middle:} Individual fit results for the ground-state energy $E_0$ in lattice units together with the result of model averaging.
\textbf{Bottom:} Model weights for the individual fits.
\label{fig:fine_pion_comparison}}
\end{figure}

The results of our procedure are shown in \cref{fig:fine_pion_comparison}.  The top pane shows the effective mass and the final result of model averaging, which is consistent with by-eye expectations.\footnote{
The effective mass of a correlation function $C(t)$ is defined by $m_{\rm eff}(t) = \log C(t) / C(t+1)$.
}
The oscillations in the effective mass are a familiar feature of staggered two-point correlation functions with nonzero momentum.
The middle pane shows intermediate results for the ground-state energy $E_0$ from individual fits.\footnote{
The ground-state energy $E_0$ comes from the first term in the spectral decomposition, Eq.~\ref{eq:2pt_spectral_decomp}.
Incidental technical complications related to staggered fermions and oscillating states are discussed in Ref.~\cite{Bazavov:2018kjg}, where this data was originally used (cf. their Eq. 2.6).
}
The green band indicates the model-averaged result for the points shown.
For clarity of presentation, the results in the middle pane are from fits with $(1+1)$ states only, i.e., 1 decaying state and 1 oscillating state.
We also tried fits including $(2+2)$, $(3+3)$, and $(4+4)$ states. 
The only qualitative difference from including more states is that good fits are obtained for smaller $t_{\rm min}$.
The model average is unchanged.
The bottom pane shows the model weights for the fits with $(1+1)$ states.
As expected, the weights peak in the middle and taper off at both ends.
When $t_{\rm min}$ is small, the fit quality rapidly declines due to contributions from excited states.
When $t_{\rm min}$ is large, \cref{eq:model_avg_cut} disfavors cutting too aggressively.
In the intermediate region, the model weights fluctuate visibly with respect to $t_{\rm tmin}$.
This behavior is related to the fact to the that model weights of Eq.~(\ref{eq:model_avg_cut}) exhibit discrete jumps as $N_{\rm cut} \in \mathbb{Z}$ is varied.
Although the total $\chi^2$ is also expected to change by roughly one unit when a degree of freedom is removed, the precise value of course depends on the details of the data.
Given the form of Eq.~(\ref{eq:model_avg_cut}), there is no  reason to expect that model weights should be a smooth function of $N_{\rm cut}$ for a finite data sample.
Overall, the model-averaged result agrees with intermediate results that went into the analysis of Ref.~\cite{Bazavov:2018kjg} to better than $1 \sigma$ \cite{EGamiz:2020}.

\subsubsection{Matrix elements from three-point correlation functions}

Model averaging also shows promising results for extraction of more complicated matrix elements.
In this example, we test the extraction of a $K\to\pi$ transition matrix $\langle \pi|J|K \rangle$, again using correlation functions that were a part of published work by the Fermilab Lattice and MILC collaborations~\cite{Bazavov:2018kjg}.
In this case, the underlying gauge-field ensemble has a lattice spacing of about $0.12$ fm, with pion and kaon masses of about $220$ and $515$ MeV, respectively.
Staggered fermions were also used for these correlators.
The calculation occurs in the rest frame of the kaon.
The pion momentum has been adjusted to be near the point of zero recoil, $q^2 \equiv (p_K - p_\pi)^2 \approx 0$.
The methodology for extracting these matrix elements is complicated but relatively standard within the lattice community.
The desired matrix element is the result of a joint correlated fit to two- and three-point functions.
In order to visualize the result, it is standard to construct a ratio $R(t, T)$ of two- and three-point functions whose asymptotic plateau is proportional to the bare lattice matrix element.
Here $T$ denotes the location of the sink operator which couples to the kaon.
In conducting such a fit, the analyst is faced with several choices: the number of states in the pion channel $(n+n)$, the number of states in the kaon channel $(m+m)$, and the fit window $t\in[t_{\rm min}, T-t_{\rm min}]$.
We refer the reader to Ref.~\cite{Bazavov:2018kjg} for additional details about fits like these.

\begin{figure}[th]
\centering
\includegraphics[height=0.75\textheight]{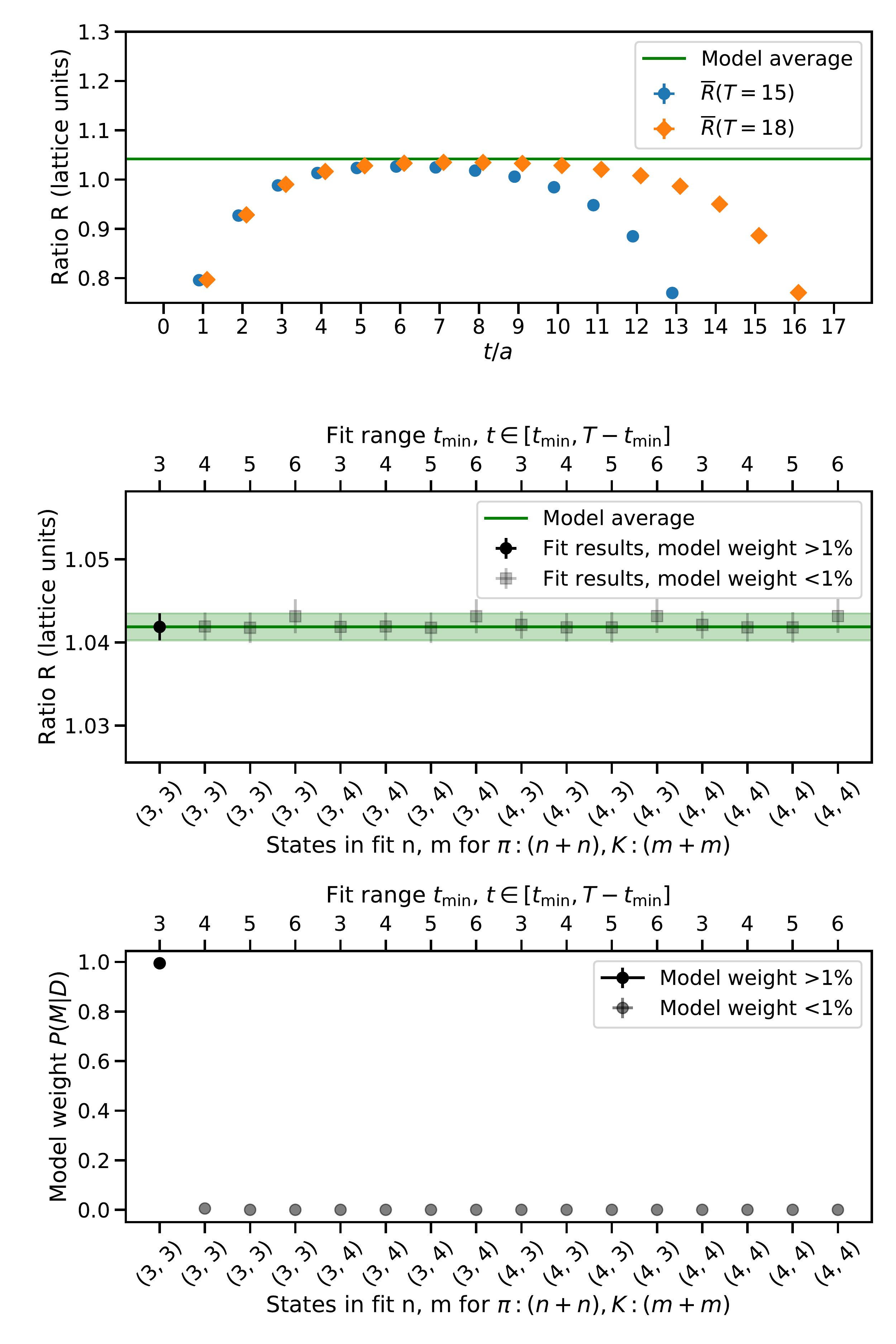}
\caption{
Model averaging results for a matrix element associated with a $K\to\pi$ transition form factor.
\textbf{Top:} A ratio of two- and three-point correlation functions (whose plateau is proportional to the matrix element $\langle \pi|J|K \rangle$) and the final result of model averaging.
\textbf{Middle:} Individual fits together with the result of model averaging.
\textbf{Bottom:} Model weights for the individual fits.
\label{fig:coarse_k2pi_comparison}}
\end{figure}

\cref{fig:coarse_k2pi_comparison} shows the result of model averaging for the matrix element. 
We adopt a flat prior model weight $\pr(M) = C$ for all choices of $(m+m)$ and $(n+n)$, which drops out of the model average.
The top pane shows the ratio $R(t,T)$ for two different sink locations alongside the result of model averaging.
The middle pane shows intermediate fit results and the model average.
The particular choices made for each of the fits is displayed along the horizontal axes, with the fit window displayed on top and the number of states on the bottom.
For instance, the leftmost point used $(n+n)=(3+3)$ states for the pion channel, $(m+m)=(3+3)$ states for the kaon channel and a fit range window $t\in[3, T-3]$.
Finally, the bottom pane gives the model weights.
In this case, all the results displayed give consistent results, and the weight of the leftmost fit is essentially unity.
This is Occam's razor, as encoded by \cref{eq:AIC_M}, at work.
For matching results, the model with the fewest parameters and most data should be preferred.
The model-averaged result, once appropriately converted into a form factor, agrees to better than $1\sigma$ with the published result of Ref.~\cite{Bazavov:2018kjg}

This example suggests another important application of the framework we are describing.
A complete analysis of a lattice matrix element might consider a more general set of fits, e.g., with different numbers of decaying and oscillating states in each channel or with different $t_{\rm min}$ cuts for the source and sink.
Scanning over all possibilities can easily produce tens or hundreds of individual fit results. 
Finding an objective selection criterion for choosing a best-fit result can be difficult.
The model weights in \cref{eq:GAPfull} and \cref{eq:AIC_M} offer a potential solution to this problem, particularly when used in conjunction with expert knowledge and the usual careful thinking.

\section{Conclusion}
\label{sec:conc}

We have presented a Bayesian approach to the problem of model averaging.
The statistical methods we describe apply very generally, though our examples have focused on practical problems in lattice gauge theory.
The context for regression problems is rather exceptional in many lattice studies, since the models often rest on firm theoretical foundations.
For instance, multi-exponential fits to correlators are based on the spectral decomposition, which only requires the existence of a positive-definite transfer matrix.
Effective field theory governs extrapolations to the chiral, continuum, or heavy-mass limit.
If a model fails to describe the data, the simulation itself is rightfully viewed with additional scrutiny.
Hypothesis testing is typically less important than reliably extracting the values of parameters capturing the physics of interest.
When predictions from nested models (say, the NLO versus NNLO predictions from effective field theory) differ slightly, it is important to be able to produce a final number with associated statistical and systematic uncertainties.
Bayesian model averaging is an attractive approach to problems like these.

Two key practical results are the model-averaged mean and variance, \cref{eq:model_avg_mean} and \cref{eq:model_avg_variance}; 
the general result \cref{eq:model_avg} allows for averaging of arbitrary expectation values of functions of fit parameters.
These formulae rely on the model weights $\pr(M|D)$.
In general, the model weights are defined through complicated integrals.
However, analytic results are available in the Gaussian approximation, which is exact for linear least-squares fitting.
For nonlinear least-squares fitting, the approximation is expected to become increasingly good for larger data sets.

For a fixed dataset with no cuts, \cref{eq:AIC_M} is the final expression used to construct the model weights used in the examples.
Since this expression is computable just using the familiar augmented $\chi^2$ and the number of parameters in the model, it is easy to include and test in existing lattice analyses.  However, it relies on taking an asymptotic limit in the sample size, and it would be interesting to study improved estimators at finite sample size in future work.

A particularly nice application of these ideas is data subset selection, which we recast as a model variation problem.
The basic observation was to reinterpret cuts on the data as additional model parameters, leading to a model weight given by \cref{eq:model_avg_cut}.
The model averaging approach gives a straightforward way to replace the common practice of tuning such data subset cuts by hand.

Broadly speaking, perhaps the most attractive feature of Bayesian model averaging is the natural appearance of Occam's razor.
The model weights appearing in \cref{eq:AIC_M,eq:model_avg_cut} favor models which use the fewest parameters while describing the most data.  Inclusion of an asymptotic bias correction to the estimated likelihood, which yields the AIC as a model-selection criterion in the limit of large sample size, is crucial to the occurrence of this effect.

Although this technique allows the data to remove much of the subjectivity from analyses including model variations, this does not extend to the choice of the model prior weights $\pr(M)$.
In the absence of specific and strong beliefs about particular models, we advocate for the use of a flat prior, i.e.\ weighting all models equally in the $\pr(M)$.  In all of the practical examples shown in the text, this is precisely what has been done.
In particular, we emphasize that one should not attempt to impose parsimony through the model priors by overweighting models with fewer parameters; this principle (i.e.~Occam's razor) is built-in to the bias-corrected model weights as we have discussed.

An interesting direction to explore in future work would be to study improved estimators at finite sample size, rather than relying on the asymptotic result to estimate the model weights.  This will necessarily involve careful treatment of the covariance matrix terms in \cref{eq:GAPfull}.  It would also be interesting to explore direct Monte Carlo evaluation of the integral in \cref{eq:model_avg_BC}, although the bias correction should be studied carefully in the context of whatever specific approach is used.
Studying the interplay of model averaging with resampling methods such as jackknife and bootstrap, commonly used in lattice analyses, would likely be a useful extension of this work.

It is worthwhile to compare Bayesian model averaging to another commonly-used technique in lattice data analysis, the empirical Bayes method \cite{Lepage:2001ym,Morningstar:2001je,Chen:2004gp}.
In this approach, the number of fit parameters is increased until the description of the physics of interest (e.g. the ground state in a two-point correlation function) becomes stable.
Empirical (data-driven) priors are used for the additional fit parameters, which are generally treated as nuisance parameters.
The use of the empirical Bayes method provides an alternative to model averaging with fixed data as presented in \cref{sec:ex1}.  
However, data subset selection can be used in conjunction with empirical Bayes modeling, and may be particularly useful for more complicated analyses where comparison of discrete model choices is required.

Within the lattice community, the empirical Bayesian method is most widely developed and applied in the analysis of correlation functions.
The usual empirical Bayesian methods also extend gracefully to ``effective field theory fits" \cite{Schindler:2008fh}, where power-counting arguments furnish firm theoretical motivation for imposing order-unity priors on certain coefficients.
In such cases, one can simply add terms until the parameters of interest stabilize.
However, for more generic fits (e.g., when the correct power-counting scheme is not \emph{a priori} obvious), adding a large number of terms may destabilize the fit and inject undesirable noise into the results.
The generality and flexibility of model averaging makes it an appealing tool for analyzing difficult problems like these, particularly in conjunction with existing methods.  Overall, we emphasize that the empirical Bayesian and model averaging techniques are complementary.

\subsection{Practical suggestions and warnings}

Model averaging has performed well for us in many test cases.
However, as with any statistical tool, the techniques we describe should not be used blindly.
In particular, model averaging should not be used as a substitute for plotting data and fits and thinking carefully about the results~\cite{Anscombe:1973}.

A basic assumption underlying this technique is that only statistically correct results are included in the model average.  Including results for fits that fail to converge numerically, for example, will likely result in incorrect answers.  Incomplete treatment of autocorrelation effects in the data will similarly yield invalid statistical estimates and thus invalid model-averaged results.

The model weights of \cref{eq:AIC_M} and \cref{eq:model_avg_cut} are useful beyond model averaging \`{a} la \cref{eq:model_avg}.
For instance, many lattice calculations oblige the analyst to make many choices beyond $t_{\rm min}$.
In this situation, the model weights can help guide the decision about which, say, dozen fit results (out of potentially hundreds) are most promising for further investigation and scrutiny using more familiar and established techniques.  Model selection is equivalent to model averaging in the limit that a single model has very high probability of correctness; this situation can naturally emerge from the data analysis, as in the example shown in \cref{fig:coarse_k2pi_comparison}.

Model averaging may be especially useful in the context of fitting models that contain discrete degrees of freedom that are not amenable to standard numerical minimization procedures.  For example, a multi-exponential model $\sum_i^\infty A_i e^{-E_i t}$ in which the sign of the amplitudes $A_i$ is \emph{a priori} unknown could be studied with improved numerical stability by fixing the signs of all included amplitudes one by one, and then averaging together the results.

In certain cases, we have found that the systematic errors due to model truncation or variation can be significantly overestimated by more conservative methods.  Revisiting old lattice analyses which are limited by systematic errors related to model variation may be worthwhile.  On the other hand, in our tests we have found excellent agreement between correlator fits with few states and those with many states; the combination of model averaging with few-state fits as a method could reduce problems related to numerical convergence and reduce the computational cost of fitting.

\begin{acknowledgments}

We thank the members of the Fermilab Lattice and MILC collaborations for generously sharing their correlator data with us for use in numerical tests.
We are especially grateful to Elvira G\'{a}miz for providing intermediate results and helping compare against published results.
We also thank Jozef Dudek, Andreas Kronfeld, Christoph Lehner, Taku Izubuchi, Tom DeGrand, Jason Chang, and Dan Hackett for helpful discussions and comments in various stages of this work.  This work was supported in part by the U.S. Department of Energy (DOE), Office of Science, Office of High Energy Physics, under Award Number DE-SC0010005 (ETN). Fermilab is operated by the Fermi Research Alliance, LLC under contract No. DE-AC02-07CH11359 with the United States Department of Energy. 

\end{acknowledgments}

\bibliography{bayes-model-combine}

\clearpage
\pagebreak

\appendix

\section{Calculation of asymptotic bias for model weights}
\label{sec:bias-appendix}

In this appendix we derive the asymptotic bias of the log-likelihood.
The form of the bias is well known in the statistics literature, and it appears in the Takeuchi Information Criterion (TIC), a generalization of the well-known Akaike Information Criterion (AIC).
What follows is not a tight mathematical proof, but rather an informal derivation designed to illustrate how the bias term arises.
For technical details, we refer interested readers to the extensive original literature~\cite{Takeuchi:1976, Stone:1977, Shibata:1989, KonishiKitagawa:1996}.
Our presentation follows closely the introduction of Ref.~\cite{DixonWard:2018}, which is particularly accessible.

The maximum likelihood estimator (MLE) \astar~is a consistent estimator of the asymptotic or ``true" \atrue.
However, the estimated log-likelihood function evaluated at \astar~is not a consistent estimator of the expected log-likehood function.
Roughly speaking, because the MLE maximizes the estimated log-likelihood, it tends to overshoot the population expected log-likelihood.

Consider the log-likelihood
\begin{align} \label{eq:LogLike}
\LL(x;\mathbf{a}) 
= \log \prod \limits_{i=1}^N L(x_i;\mathbf{a})
= \sum\limits_{i=1}^{N} \log L(x_i;\mathbf{a}),
\end{align}
where $L(x_i;\mathbf{a})$ is the likelihood for a single sample $x_i$ evaluated with model parameters $\mathbf{a}$.
The sample and population expectation values of a function $g(x)$ are defined according to \begin{align}
\En [ g(x) ] 
	&= \frac{1}{N}\sum\limits_{i=1}^{N} g(x_i)
	\equiv \avgn{g(x)} \label{eq:SampleExpectation}\\
\Ez [ g(z) ] 
	&= \int dz\, f(z) g(z)
	\equiv \avgz{g(z)},
\end{align}
where $f(z)$ is the population distribution from which the samples $\{x_i\}$ are presumed to be drawn.
Because \astar~and \atrue~maximize their respective log-likelihoods, they are solutions to the usual equations:
\begin{align}
    \left.
        \avgn{\partial_\mathbf{a} \LL(x;\mathbf{a})} 
    \right\vert_{\mathbf{a}=\astar}
    &=0 \label{eq:astar_def}\\
    \left.
        \avgz{\partial_\mathbf{a} \LL(z;\mathbf{a})}
    \right\vert_{\mathbf{a}=\atrue}
     &=0
    \label{eq:atrue_def}.
\end{align}
Note that, for a fixed number of samples $N$, $\astar$ is a fixed number.
The sample Fisher information matrix $I_N$ and the negative sample Hessian matrix $J_N$ are defined as
\begin{align}
I_{N,xy}(\mathbf{a})
    &\equiv \frac{1}{(N-1)} \sum_{i=1}^N
    \left(\frac{\partial \LL(x_i;\mathbf{a})}{\partial a_x}\right)
    \left(\frac{\partial \LL(x_i;\mathbf{a})}{\partial a_y}\right)\\
    &= \frac{1}{4(N-1)} \sum_{i=1}^N \left(\frac{\partial \chi_{i}^2}{\partial a_x}\right) \left(\frac{\partial \chi_i^2}{\partial a_y}\right), \\
J_{N,xy}(\mathbf{a})
    &\equiv
    -\frac{1}{N} \sum_{i=1}^N 
    \frac{\partial^2 \LL(x_i|\mathbf{a})}{\partial a_x \partial a_y}
    = \frac{1}{2N} \sum_{i=1}^N 
    \frac{\partial^2 \chi_{i}^2}{\partial a_x \partial a_y}.
\end{align}
The final equalities are valid for the special case of least-square fitting, where $-2 \log L(x;\mathbf{a}) = \chi^2$ (cf. \cref{sec:lsq} for additional notation).
Similarly, 
\begin{align}
I_{xy}(\mathbf{a})
    &\equiv\Ez \left[
    \frac{\partial \log L(z;\mathbf{a})}{\partial \mathbf{a}_x} 
    \frac{\partial \log L(z;\mathbf{a})    }{\partial \mathbf{a}_y}
    \right]\\ 
J_{xy}(\mathbf{a})
    &\equiv -\Ez \left[
    \frac{\partial^2 \log L(z;\mathbf{a})}{\partial \mathbf{a}_x \partial \mathbf{a}_y}
    \right].
\end{align}    
    
With~\cref{eq:LogLike} and~\cref{eq:SampleExpectation}, the total log-likelihood can be written as $\LL(x;\mathbf{a})= N \avgn{\LL(x;\mathbf{a})}$.
The bias in the log-likelihood is defined as the difference between its estimated and expected values,
\begin{align}\label{eq:bias}
b(\astar(x)) \equiv N \avgn{ \LL(x;\astar(x)) - \avgz{ \LL(z;\astar(x))}}.
\end{align}
We are interested in the behavior of this bias in the limit of many samples, $N\to\infty$.
To emphasize the dependence on the data, we have written $\astar = \astar (x)$.
To evaluate the bias explicitly and resolve the mixed expectation value, it helps to add and subtract terms judiciously:
\begin{align}\label{eq:BiasRewrite}
\begin{split}
b(\astar(x)) =
	&N\Big( \avgn{\LL(x;\astar(x))} - \avgn{\LL(x;\atrue)} \Big) \\
	&+N\Big( \avgn{\LL(x;\atrue)} - \avgz{\LL(z;\atrue)} \Big) \\
	&+N\Big( \avgz{\LL(z;\atrue)} - \avgn{\avgz{\LL(z;\astar(x))}} \Big).
\end{split}
\end{align}
This trivial rewriting pays immediate dividends.
The first and third terms involve matching expectation values at nearby points and are amenable to Taylor expansion.
As we will argue shortly, the second term vanishes.

Before evaluating each term, we quote a useful technical result due to White~\cite{White1982}.
The necessary regularity conditions for this result, which White calls the ``usual maximum likelihood regularity conditions,'' are stated carefully and at length in Ref.~\cite{White1982}.

\begin{theorem}\label{thm:asymptotic_normality}
\textbf{Asymptotic Normality.}
Given White's regularity conditions, the difference $(\astar - \atrue)$ is asymptotically normally distributed with mean zero and width $C(\atrue)$,
\begin{align}
\sqrt{N} (\astar - \atrue) \stackrel{N\rightarrow\infty}{\sim} \Normal(0, C(\atrue)),
\end{align}
where 
\begin{align}
C(\atrue) &= J^{-1}(\atrue) I(\atrue) J^{-1}(\atrue)\\
C_N(\astar) &= J_N^{-1}(\astar) I_N(\astar) J_N^{-1}(\astar)
\end{align}
is a product of (inverse) Hessian and Fisher matrices.
Moreover, $C_N(\astar) \stackrel{N\to\infty}{\rightarrow} C(\atrue)$ element by element.
\end{theorem}

Now we turn to the evaluation of~\cref{eq:BiasRewrite}, beginning with the first term.
Expanding around the MLE point \astar~gives
\begin{align}
N\left(\avgn{\LL(x;\astar(x)} - \avgn{\LL(x;\atrue)}\right)
&= -\frac{N}{2} \avgn{ (\atrue - \astar(x)) \frac{\partial^2 \LL}{\partial \mathbf{a} \partial \mathbf{a}^\prime} (\atrue - \astar(x)) } \\
&\stackrel{N\to\infty}{\longrightarrow} + \frac{1}{2} \tr \left[ I(\atrue) J^{-1}(\atrue) \right].
\end{align}
In the first equality, the linear term vanishes by the definition of \astar, \cref{eq:astar_def}.
The second line follows from the asymptotic normality of $\sqrt{N} (\astar - \atrue)$ and the fact that $-\frac{\partial^2 \LL}{\partial \mathbf{a} \partial \mathbf{a}^\prime}$ converges to $J$ in probability.
The final line also uses a standard result about expectation values of random quadratic forms, 
\begin{align}
\mathbb{E}[\varepsilon^T A \varepsilon] 
= \tr [\Sigma A] + \mu^T A \mu,
\end{align}
where $\varepsilon$ is a random variable with mean $\mu$ and covariance $\Sigma$.

The second term in~\cref{eq:BiasRewrite} vanishes.
To see this, first observe that both terms are evaluated at the same fixed parameters \atrue, which remain unchanged by the limit $N\to\infty$.
Next, note that the estimated log-likelihood converges point-by-point to the asymptotic distribution.
Therefore, the difference vanishes in this limit.
This argument fails when the MLE $\astar(x)$ is involved, since $\astar(x)$ depends on the data and thus itself moves with $N$.

Finally, we consider the third term, which contains population expectation values.
In this case it is useful to expand around \atrue, since the linear term will vanish by \cref{eq:atrue_def}:
\begin{align}
N &\left(\avgz{\LL(z;\atrue)} - \avgn{\avgz{\LL(z;\astar(x))}} \right)\\
    &=  N\left(\avgz{\LL(z;\atrue)}
        - \avgn{\avgz{\LL(z;\atrue)}}
        + \frac{1}{2}\avgn{(\astar-\atrue)J(\atrue)(\astar-\atrue)} \right)\\
    &=  N\left(\avgz{\LL(z;\atrue)}
        - \avgz{\LL(z;\atrue)}
        + \frac{1}{2}\avgn{(\astar-\atrue)J(\atrue)(\astar-\atrue)}\right) \\
    &=  \frac{N}{2}\avgn{(\astar-\atrue)J(\atrue)(\astar-\atrue)} \\
    &\stackrel{N\to\infty}{\longrightarrow}
        \frac{1}{2} \tr \left[ I(\atrue) J^{-1}(\atrue) \right].
\end{align}
The final line follows from the asymptotic normality of $\sqrt{N} (\astar - \atrue)$ and the formula for random quadratic forms.
Combining results for all three terms, we see that 
\begin{align}
b(\astar(x))
&\stackrel{N\to\infty}{\longrightarrow}
    +\frac{1}{2} \tr \left[ I(\atrue) J^{-1}(\atrue) \right]
    + 0
    + \frac{1}{2} \tr \left[ I(\atrue) J^{-1}(\atrue) \right]\\
&= \tr \left[ I(\atrue) J^{-1}(\atrue) \right] 
\end{align}
As indicated, this result is evaluated at the (unknown) parameters \atrue.
However, since the sample $I_N(\astar)$ and $J_N(\astar)$ are consistent estimators of $I(\atrue)$ and $J(\atrue)$, the bias may be evaluated using them instead~\cite{Shibata:1989, KonishiKitagawa:1996, DixonWard:2018}.

For the sake of concreteness, the proof sketched here has used the MLE~$\astar$.
However, a similar bias term is expected to be present quite generally.
For instance, Theorem 2.1 of Ref.~\cite{KonishiKitagawa:1996} proves the existence of bias for a more general class of estimators.
Roughly speaking, the bias arises from finite-sample-size fluctuations in the data and not from the choice of the maximum likelihood estimator itself.  Due to the generality of this bias term, we include the correction in the general formula \cref{eq:model_avg_BC} and not only in the following Gaussian approximation.

So far the discussion has been for general log-likelihoods.
Now we specialize to the case of least-square fitting, where $-2\log L(x;\mathbf{a})= \chi^2(x; \mathbf{a})$.
Taking $\chi^2(x; \astar) \mapsto \chi^2(x; \astar) + 2\,\tr[I_N(\astar) J_N^{-1}(\astar)]$ as in \cref{eq:GAPfull} removes this bias.

In most cases of interest in lattice gauge theory (spectral decomposition of correlation functions, effective theory descriptions of the chiral-continuum limit, etc.), the correct model for the data is assumed to be known.
When the model is specified correctly, the following theorem, proven by White in Ref.~\cite{White1982} with careful attention to regularity conditions, gives the familiar equivalence between the negative Hessian and the Fisher information matrix.
\begin{theorem}\label{thm:information_matrix_equivalence}
\textbf{Information Matrix Equivalence.}
Given White's regularity conditions and assuming that the model is specified correctly,
\begin{align} \label{eq:information_matrix_equivalence}
C(\atrue) = J(\atrue)^{-1} = I(\atrue)^{-1}. 
\end{align}
In other words, the negative Hessian equals the Fisher information matrix.
Likewise, if the model is specified incorrectly, then \cref{eq:information_matrix_equivalence} fails to hold in general, i.e., $J(\atrue)$ is generically not equal to $I(\atrue)$.
\end{theorem}

For completeness, we sketch how this equivalence arises and how it can break down. Consider the log-likelihood $\log L(z;\mathbf{a})$.
Elementary use of the product rule shows that
\begin{align}\label{eq:LLderivatives}
\frac{\partial^2 \log L(z;\mathbf{a})}{\partial \mathbf{a}_x \partial \mathbf{a}_y}  
    = \frac{1}{L(z;\mathbf{a})}
    \frac{\partial^2  L(z;\mathbf{a})}{\partial \mathbf{a}_x\partial \mathbf{a}_y} 
    - \frac{\partial \log L(z;\mathbf{a})}{\partial \mathbf{a}_x} 
    \frac{\partial \log L(z;\mathbf{a})    }{\partial \mathbf{a}_y} 
\end{align}
Taking the expectation of both sides (and using the likelihood function as the probability distribution function) gives:
\begin{align}\label{eq:HessianFisher}
    \Ez \left[
    \frac{\partial^2 \log L(z;\mathbf{a})}{\partial \mathbf{a}_x \partial \mathbf{a}_y}
    \right]
    =
    -\Ez \left[
    \frac{\partial \log L(z;\mathbf{a})}{\partial \mathbf{a}_x} 
    \frac{\partial \log L(z;\mathbf{a})    }{\partial \mathbf{a}_y}
    \right],
\end{align}
since the first term on the right-hand side vanishes:
\begin{align}
    \Ez \left[
    \frac{1}{L(z;\mathbf{a})}
    \frac{\partial^2  L(z;\mathbf{a})}{\partial \mathbf{a}_x\partial \mathbf{a}_y} 
    \right]
    &\equiv 
    \int dz L(z;\mathbf{a}) 
    \left(
    \frac{1}{L(z;\mathbf{a})}
    \frac{\partial^2  L(z;\mathbf{a})}{\partial \mathbf{a}_x\partial \mathbf{a}_y}
    \right)\\
    &=\frac{\partial^2}{\partial \mathbf{a}_x\partial \mathbf{a}_y}
    \int dz L(z;\mathbf{a}) 
    = \frac{\partial^2}{\partial \mathbf{a}_x\partial \mathbf{a}_y} 1 = 0.
\end{align}
\cref{eq:HessianFisher} is the familiar equivalence between the negative Hessian and the Fisher matrix.
The vanishing expectation value of the first term on the right-hand side of~\cref{eq:LLderivatives} depends on the appearance of $\LL(x;\mathbf{a})$ in the numerator.
That is, it assumes that the true likelihood appears within the model space.
In general, if the model is incorrectly specified, \cref{thm:information_matrix_equivalence} says that the equivalence between the negative Hessian and Fisher matrices ceases to hold.
White also provides specification-robust procedures~\cite{White1982}.
We emphasize that many statistical analyses in lattice gauge theory are expected to be in the privileged position of knowing the correct model.

\section{Equivalence of sample-based and mean-based forms of chi-squared likelihood}
\label{sec:chi-squared-appendix}

In this brief appendix, we discuss the definition of the likelihood function in \cref{eq:Lsample} and \cref{eq:chi2sample}, based on the individual sample values $\chi_i^2$.  This definition is not manifestly equivalent to the standard definition based on the data means,
\beq \label{eq:chi2mean}
\chi^2 \equiv (\bar{y} - f_M(\mathbf{a}))^T (\Sigma/N)^{-1} (\bar{y} - f_M(\mathbf{a})),
\eeq
where $\bar{y} = \frac{1}{N} \sum_{i=1}^N y_i$ is the sample mean (cf. \cref{sec:lsq} for additional notation).
However, it is straightforward to prove that the two definitions agree up to an additive constant,
\beq
\sum_{i=1}^N \chi_i^2 = (N-1)d + \chi^2.
\eeq
where $d$ is the dimension of the observation vector $y_i$, as defined in the main text.  The proof follows by elementary matrix manipulations.
By definition, the left-hand side is
\begin{align}
\sum_{i=1}^N \chi_i^2 
&=\sum_{i=1}^N (y_i - f_M(\mathbf{a}))^T \Sigma^{-1} (y_i - f_M(\mathbf{a}))\\
&=\sum_{i=1}^N (y_i^T) \Sigma^{-1} y_i - 2 N f_M(\mathbf{a})^T \Sigma^{-1} \bar{y} + N f_M(\mathbf{a})^T \Sigma^{-1} f_M(\mathbf{a})\\
&=(N-1)d + N \bar{y}^T \Sigma^{-1} \bar{y} - 2 N f_M(\mathbf{a})^T\Sigma^{-1} \bar{y} + N f_M(\mathbf{a})^T \Sigma^{-1} f_M(\mathbf{a}) \\
&=(N-1)d + (\bar{y} - f_M(\mathbf{a}))^T \left(\Sigma/N\right)^{-1} (\bar{y} - f_M(\mathbf{a})) \\
&\equiv (N-1)d + \chi^2,
\end{align}
where we have used the definition of the sample mean, $\bar{y} = \frac{1}{N} \sum_{i=1}^N y_i$.
To obtain the third equality, we have used the relation $\sum_{i=1}^N y_i y_i^T = (N-1) \Sigma + N \bar{y} \bar{y}^T$ (which follows from rearrangement of the definition of the sample covariance matrix), 
as well as the matrix identity $\sum_{i=1}^N y_i^T \Sigma^{-1} y_i = \sum_{i=1}^N \tr[\Sigma^{-1} y_i y_i^T]$.

This proof shows clearly how the standard error of the mean, $\sigma_{\bar{y}}^2 = \sigma^2 / N$, arises in $\chi^2$ from the sample covariance matrix.  We also clearly see that minimization of chi-squared to find the best-fit parameters $\mathbf{a}^\star$ will give identical results in either case.

\section{Linear and Nonlinear Models}\label{sec:linear_vs_nonlinear}
Both linear and nonlinear models appear commonly in lattice gauge theory analyses.
Linear models are linear functions of the fit parameters.
A common example of the linear case is an order-$p$ polynomial model, which can be written as
\begin{align}
    f(\mathbf{x}; \mathbf{a})
    &= X \mathbf{a}\\
    \begin{pmatrix}
    f_0\\   f_1\\   \vdots\\    f_d
    \end{pmatrix}
    &= \begin{pmatrix}
    0       & x_0       & x_0^2     & \cdots    & x_0^p \\
    0       & x_1       & x_1^2     & \cdots    & x_1^p \\
    \vdots  & \vdots    & \vdots    &           & \vdots\\
    0       & x_d       & x_d^2     & \cdots    & x_d^p \\
    \end{pmatrix}
    \begin{pmatrix}
    a_0\\   a_1\\   \vdots\\    a_p
    \end{pmatrix}
\end{align}
where $\mathbf{x}$ is the $d$-dimensional data vector, $\mathbf{a}$ is the $p$-dimensional vector of model parameters, and X is the $(d\times p)$ design matrix.
We note that design matrix is rectangular but not generally square.
In the special case of one dimensional data, the previous equation reduces to the familiar $f(x;\mathbf{a}) = a_0 + a_1 x + a_2 x^2 + \dots + a_p x^p$.

Nonlinear models are nonlinear functions of the fit parameters.
The most important example in lattice gauge theory is the spectral decomposition of Euclidean two-point correlation functions, for which a $p$-state model takes the form
\begin{align}\label{eq:2pt_spectral_decomp}
f(t;\mathbf{a}) = A_0 e^{-E_0 t} + A_1 e^{-E_1 t} + \cdots + A_p e^{-E_p t},
\end{align}
where here $\mathbf{a}$ denotes the full set of $2p$ model parameters,
\begin{align}
\mathbf{a} = \{A_0, A_1, \cdots, A_p\}\cup\{E_0, E_1, \cdots, E_p\}.
\end{align}
Since each term involves the product of an amplitude $A_i$ with $e^{-E_i t}$, the model function is clearly nonlinear in the fit parameters.

We emphasize that the distinction between linear and nonlinear models applies only to the functional dependence on the model parameters.
As the polynomial example shows, a generic linear model can be an arbitrary nonlinear function of the data.

\end{document}